\documentclass[12pt,a4paper]{article}
\pdfoutput=1

\usepackage{array}
\usepackage{color}
\usepackage{latexsym}
\usepackage{amsthm}
\usepackage{amsmath}
\usepackage{amssymb}
\usepackage{mathrsfs}
\usepackage{multirow}
\usepackage[final]{pdfpages}
\usepackage{rotating}
\usepackage{textcomp}
\usepackage{url}
\usepackage{verbatim}
\usepackage{wrapfig}
\usepackage{amsfonts}
\usepackage{comment}
\usepackage{epigraph}
\usepackage{anysize}
\usepackage{bold-extra}
\usepackage{color}
\usepackage{enumerate}
\usepackage{fancyhdr}
\usepackage{graphicx}
\usepackage[table]{colortbl}
\usepackage[utf8]{inputenc}
\usepackage{cancel}
\usepackage{dsfont}
\usepackage{epsfig}
\usepackage{slashed}
\usepackage{bbold}
\usepackage{psfrag}
\usepackage[svgnames]{xcolor}
\PassOptionsToPackage{caption=false}{subfig}
\usepackage{subcaption}
\usepackage{xfrac}
\usepackage{multirow}
\usepackage{booktabs}
\usepackage[colorlinks=true,linkcolor=MediumBlue,citecolor=Green,urlcolor=violet]{hyperref}
\usepackage{cite}
\usepackage[normalem]{ulem}
\usepackage{float} 
\usepackage[bottom]{footmisc} 
\usepackage{bbm}
\usepackage{cancel}

\usepackage{tikz}

\newcommand{\eq}[1]{Eq.~(\ref{#1})}

\newcommand{\noindentFR}[1]{\vspace{5mm}\noindent {\bf #1}\\}

\def\nn{\nonumber}
\def\bal#1\eal{\begin{align}#1\end{align}}

\def\bel#1{\begin{equation} \label{#1}}

\newcommand{\be}{\begin{equation}}
\newcommand{\ee}{\end{equation}}
\newcommand{\beq}{\begin{equation}}
\newcommand{\eeq}{\end{equation}}
\newcommand{\bea}{\begin{eqnarray}}
\newcommand{\eea}{\end{eqnarray}}

\newcommand{\LL}{\mathcal{L}}

\newcommand{\SM}{\textrm{SM}}

\usepackage{mathrsfs}

\def\nn{\nonumber}

\newcommand{\cmrule}{\midrule[0.25mm]}
\newcommand{\ctoprule}{\toprule[0.5mm]}
\newcommand{\cbottomrule}{\bottomrule[0.5mm]}

\newcommand{\cc}{\cellcolor[rgb]{0.9,0.9,0.9}}

\newcommand{\BSM}{\mathrm{BSM}}

 \def\b{\beta}

 \def\Ph{\Phi}

\def\mO{\mathcal{O}}

\newcommand{\ba}{\begin{eqnarray}}
\newcommand{\ea}{\end{eqnarray}}

\begin{document}


\begin{flushright}
\hspace{3cm} 
LHC-HXSWG-2019-006
\end{flushright}
\vspace{.6cm}
\begin{center}

{\LARGE \bf  BSM Benchmarks for Effective Field Theories\\ \vspace{0.5cm} in Higgs and Electroweak Physics}\\[0.5cm]

\vspace{1cm}
{
{D.~Marzocca$^{a}$, 
F.~Riva$^{b}$ }(Editors),
J. Criado$^c$,
S.~Dawson$^d$,
J. de Blas$^{e,f,g}$, 
B.~Henning$^{b}$,\\[2mm]
D.~Liu$^{h}$,
C.~Murphy$^d$,
M. Perez-Victoria$^c$, 
J. Santiago$^c$,
L.~Vecchi$^{i}$, Lian-Tao Wang$^{j}$
}
\\[7mm]
 {\it \small

$^a$ INFN Sezione di Trieste, SISSA, via Bonomea 265, 34136 Trieste, Italy\\[0.1cm]
$^b$ D\'epartment de Physique Th\'eorique, Universit\'e de Gen\`eve, 24 quai Ernest-Ansermet, 1211 Gen\`eve 4, Switzerland\\[0.1cm]
$^c$CAFPE and Departamento de Fisica Teorica y del Cosmos, Universidad de Granada, Campus de Fuentenueva, E-18071, Granada, Spain\\[0.1cm]
$^d$Department of Physics, Brookhaven National Laboratory, Upton, N.Y., 11973, U.S.A.\\[0.1cm]
$^{e}$Dipartimento di Fisica e Astronomia Galileo Galilei, Universit\'a di Padova, Via Marzolo 8, I-35131 Padova, Italy\\[0.1cm]
$^{f}$INFN, Sezione di Padova, Via Marzolo 8, I-35131 Padova, Italy\\[0.1cm]
$^{g}$Institute of Particle Physics Phenomenology, Durham University, Durham DH1 3LE, UK \\[0.1cm]
$^{h}$High Energy Physics Division, Argonne National Laboratory, Lemont, IL 60439, USA\\[0.1cm]
$^{i}$Theoretical Particle Physics Laboratory, Institute of Physics, EPFL, Lausanne, Switzerland \\[0.1cm]
$^{j}$ Department of Physics, Enrico Fermi Institute, and Kavli Institute for Cosmological Physics,University of Chicago, 5640 S Ellis Ave, Chicago, IL 60637, USA
 }

\end{center}

\bigskip \bigskip \bigskip

\centerline{\bf Abstract} 
\begin{quote}
Effective Field Theories (EFTs) capture effects from heavy dynamics at low energy and represent an essential ingredient in the context of Standard Model (SM) precision tests. This document gathers a number of relevant scenarios for heavy physics beyond the SM and presents explicit expressions for the Wilson coefficients in their low-energy EFT. It includes \emph{i)} weakly coupled scenarios in which one or a few particles of different spins and quantum numbers interact linearly with the SM and generate EFT effects at tree-level, \emph{ii)}  scenarios where heavy particles interact quadratically whereupon the resulting EFT arises only at loop-level and \emph{iii)} strongly coupled scenarios where the size of Wilson coefficients is controlled by symmetry arguments.
This review aims at motivating experimental EFT studies in which only a subset of all possible EFT interactions is used, as well as facilitating the theoretical interpretation of EFT fits.
\end{quote}

\newpage

\tableofcontents

\section{Introduction}

Standard Model (SM) precision tests are gradually gaining importance as a search tool for physics beyond the SM (BSM). Indeed, heavy resonances beyond direct collider reach still impact SM processes, in a way that can be searched for via precision tests. Effective field theories (EFTs) provide a way to parametrise  these effects and translate between the language of BSM dynamics and the language of deviations from the SM \cite{deFlorian:2016spz}.

One of the often emphasized features of EFTs  is their model-independence. One  can stop bothering about the plethora of microscopic BSM models and focus on a  number of the most relevant EFT operators: any BSM model can be matched to these, so that an analysis in terms of these can be matched to any BSM model. This has led to the development of global fits to test simultaneously all possible (leading) operators against all possible observables, with the noble goal of providing, once and for all, a unique and versatile map from experiment to theory.

While absolute model-independence remains an ideal goal, in practice it can often hinder the ability to exploit the full power of the EFT approach, thereby limiting what we can learn about BSM models.
First of all, the number of operators at the leading (dimension-6) order is already overwhelming \cite{Alonso:2013hga}, so that restrictive assumptions appear in most analyses. These include  Minimal Flavour Violation \cite{DAmbrosio:2002vsn}, or a focus on universal theories (where BSM couples only to bosons) \cite{Barbieri:2004qk,Englert:2019zmt}, etc. Moreover, the inclusion of many operators is inefficient, as different ones can contribute in equal (or similar) ways to a given observable. This leads to (nearly) flat directions, i.e. combination of  Wilson coefficients that do not contribute to an observable at tree-level.
When presented in the context of a global fit as marginalised results, these directions give the false impression that the experimental reach is poorer than it actually is. {Such flat directions also raise the issue of how to present results of global fits, since the Gaussian approximation is often not appropriate in these cases. Moreover, it is often the case that some of the operators relevant for such flat directions are more suppressed than others in a wide class of BSM scenarios.}
Model-independence is also a burden when discussing the question of EFT validity, for which specific assumptions about the underlying dynamics are necessary to properly phrase the question.
Finally, one gets valuable insight into the SM by testing it against specific hypothesis; insight which may not be apparent in a generic context.

In this document we therefore provide a different point of view on the usage of the SM EFT: we gather a number of BSM \emph{benchmark} scenarios that match to EFTs with a finite (and controllable) number of operators.
This should not to be thought of as picking random points in {the full EFT} 
parameter space, but rather as identifying the (large) regions that are best motivated from some plausible underlying dynamics. This exercise---which produces a rich map between certain BSM models and the coefficients of EFT operators---is certainly useful for theorists to directly interpret experimental results in terms of model parameters. Moreover, it has utility for experimentalists, who can use it to focus their analysis on a small number of operators as well as to motivate more sharply specific searches.

Given the countless directions in which BSM physics has developed, it is impossible (as well as pointless) to provide a complete catalog. Here we identify two regimes in which BSM dynamics can be classified: at weak and strong coupling.

In the first category we study simple scenarios in which the SM is supplemented with a minimal number of additional, weakly coupled degrees of freedom. These can be thought of as the lightest states of more complex scenarios (also known as simplified models).
For example, new  resonances $\Phi$ can couple linearly to some operator in the SM $\mathcal{O}_{\textrm{SM}}$, $\LL \supset g\Phi\mathcal{O}_{\textrm{SM}}$, where $g$ is a new coupling. Their tree-level exchange can then generate sizeable effective operators of the form $g^2\mathcal{O}_{\textrm{SM}}^2/M_{\Phi}^2$, with $M_{\Phi}$ the mass of the new resonance. Sections \ref{sec:minimalext} and \ref{sec:ESS_SMEFT} classify  possible resonances $\Phi$, with different quantum numbers, which give rise at tree-level to dimension-6 operators. 

It is also plausible that the new physics does not couple linearly to the SM, but rather quadratically as $\LL \sim\Phi^2\mathcal{O}_{\textrm{SM}}$, due for instance to accidental symmetries and conserved quantum numbers in the BSM sector, such as $R$-parity in supersymmetric models or $T$-parity in Little-Higgs models. Then, the leading BSM imprints on SM processes necessarily appear at loop-level. We discuss some scenarios of this class in section \ref{sec:loop} and identify interesting patterns that can be explicitly tested. For instance, it is found that the ratio between some dimension-6 effects that appear in Drell-Yan processes and those that appear in anomalous triple gauge couplings (TGC) is uniquely determined by the scalar/fermion/vector nature of the particle that is being integrated out.

In the strongly  coupled regime, instead, this perturbative classification loses its meaning: loops of new physics can become sizeable and there might be a multitude of resonances with different quantum numbers appearing {at the same mass scale.} There are still situations, though, where we can reliably build a well-structured EFT. This happens when the strong sector is endowed with (non-linearly realised) symmetries that imply selection and power-counting rules at low-energy. The most notable example is Composite Higgs (CH) models, discussed in section \ref{sec:CompositeHiggs}, where the Higgs is in fact a pseudo-Nambu-Goldstone boson of a BSM global symmetry \cite{Kaplan:1983fs,Kaplan:1983sm,Contino:2003ve,Giudice:2007fh}. Another example concerns instead multipolar interactions \cite{Liu:2016idz}, a sort of vector-compositeness, discussed in section \ref{sec:remedios}.

The main results of this document are summarized in tables \ref{tab:bottom-up} (weakly coupled tree-level BSM), \ref{tab:dim6ops_note} (weakly coupled scalars), \ref{tbl:MSSM}-\ref{tbl:EW_doublet} (loop-level), and tables \ref{tab2} and \ref{tab:powercounting} for strongly coupled models (Higgs and vector compositeness). 
These results can be used in many ways. An experimental physicist could consult these tables to see whether a particular choice of operators, motivated by the experimental sensitivity of her/his experiment, can be generated in any of these models, and these could be used to motivate a given analysis or justify a limited choice of operators. For instance, di-boson processes are typically analysed in terms of operators like $\mathcal{O}_{3W}$ (see table~\ref{tab1}), which, from tables  \ref{tbl:MSSM} and \ref{tab:powercounting}, we learn can be generated at loop level by light stops, or in strongly coupled models of vector-compositeness.
Furthermore table \ref{tab1} provides a partial list of how different operators enter into different process, that could be used as a first tool of orientation.

A theoretical physicist might instead be interested in how exactly a constraint on a given operator translates in terms of the physical parameters in a given model, something that can be directly read from the above tables. For instance from table  \ref{tab:bottom-up} we learn that a per-mille level constraint on the $T$-parameter (operator $\mathcal{O}_{\phi D}/m_W^2$ in the notation of table \ref{tab:bottom-up}) puts a bound on the mass of a neutral vector singlet  of order 3.5 TeV (for O(1) couplings, using the explicit expression in table \ref{tab:vsinglet}).

\subsection{The SM Effective Field Theory}

Effective field theories are useful in UV theories where it is possible to make a distinction between {\bf light and heavy} degrees of freedom. The heavy states have masses much larger than the typical momentum scale characterizing the processes under investigation, and can thus be ``integrated out" leaving a Lagrangian for the light degrees of freedom that can be expressed in terms of local operators. In the SM EFT the light degrees of freedom are the SM fields -- this includes the Higgs boson as part of an {\bf electroweak doublet}.\footnote{This assumption, though not globally accepted in the literature, is motivated by the apparent lack of other sources of electroweak symmetry breaking, as expressed by the remarkable agreement between the SM and precision electroweak data, as well as Higgs coupling measurements. A more general parametrisation is provided by the Higgs Effective Field Theory (HEFT)~\cite{deFlorian:2016spz}; a comparison between HEFT and SMEFT provides the basis to test the Higgs as a doublet hypothesis.}

In this situation, any observable involving the light degrees of freedom (here the SM), as computed in the full UV theory, can be expanded as a Taylor series {in powers of $1/\Lambda^n$, where $\Lambda$ is the mass scale of the heavy states, or the cutoff of the EFT}. 
The SM EFT includes all such possible expansions that can originate from any possible UV model and it parameterizes all  possible local interactions among SM fields in a bottom-up approach:
\begin{equation}
  \mathcal{L}_{\text{EFT}} = \mathcal{L}_{\text{SM}} + \sum_i \frac{c_i}{\Lambda^{d_i-4}}\mathcal{O}_i,
  \label{eq:L_SMEFT}
\end{equation}
where the $\mathcal{O}_i$ are gauge-invariant operators of mass dimension \(d_i\) built from SM fields and their derivatives and we have collected the dimension $\leq 4$ operators belonging to the SM inside \(\mathcal{L}_{\text{SM}}\). 
To each operator \(\mathcal{O}_i\) we associate a dimensionless coupling constant \(c_i\), called the Wilson coefficient.

In specific models, heavy states leave an imprint in the low-energy dynamics that can be captured by the effective Lagrangian \eq{eq:L_SMEFT} where the heavy states have been integrated out  and ``matched'' onto the SM EFT to obtain the resulting Wilson coefficients $c_i$ as functions of the specific model parameters.
The different coefficients are then correlated under the assumption of particular high-energy models. 

For a physical process occuring at some characteristic energy \(E\), the associated action cost of an operator roughly scales as \(\int d^4x \, \mathcal{O}_i \sim E^{d_i-4}\). For \(E \ll \Lambda\), the effective Lagrangian becomes useful because we can truncate the expansion and keep only a finite number of operators.  This document will be concerned with the leading effects in \eq{eq:L_SMEFT} which, under the assumption of approximate $B$ and $L$  symmetries, correspond to operators of dimension~6.

Field redefinitions in the EFT do not modify observable predictions, but can change the form of the Lagrangian \eq{eq:L_SMEFT}. This leads to different possibilities for chosing a minimal set of operators  $\mathcal{O}_i$.
The most popular choices include the so-called Warsaw basis~\cite{Grzadkowski:2010es} (ideal if new physics couples directly to fermions), of which we summarize the most important operators for Higgs physics in  table~\ref{tab:bottom-up}; and the SILH basis~\cite{Giudice:2007fh} (ideal for universal theories, where the new physics couples mainly to bosons).\footnote{{An extension of the SILH set of operators to a complete basis was done in \cite{Elias-Miro:2013eta}, see also \cite{Wells:2015uba}.}} The SILH basis also turns out to be  convenient  because it naturally distinguishes operators that can be generated at tree- and at loop-level in simple theories (see table~\ref{tab2} in section~\ref{sec:CompositeHiggs}). The \emph{Rosetta} tool presented in Ref.~\cite{Falkowski:2015wza} provides an automatic translation between Wilson coefficients in different bases.

Each SM process will be modified by a number of operators. While it is difficult to provide a complete map between operators and processes, table~\ref{tab1} summarises some of the main  processes that can be associated with each operator. These include on-shell processes in which a Higgs appears on-shell, and processes with longitudinally polarized vector bosons which, according to the equivalence theorem, are affected by the same operators that enter in Higgs physics, see e.g. \cite{Henning:2018kys} and references therein.

\begin{table}[h!]
\begin{center}
\resizebox{\textwidth}{!}
{
\begin{tabular}{|c|c||c|c|} 
\hline
$\mathcal{O}$ & Operator definition & Main On-shell (Higgs) & Dominant Off-shell\\
\hline 
${\cal O}_H$ & $\frac{1}{2}~\partial_\mu(H^\dagger H)\partial^\mu(H^\dagger H)$ & { $h\to\psi\bar\psi, VV^*$
}
 & { $V_LV_L\to V_LV_L,hh$}   \\
${\cal O}_T$ & $\frac{1}{2}~(H^\dagger\overleftrightarrow{D_\mu}H)(H^\dagger\overleftrightarrow{D^\mu}H)$ & { $h\to ZZ^*$ 
}
 & { $V_LV_L\to V_LV_L,hh$}   \\
${\cal O}_6$ & $\lambda_h(H^\dagger H)^3$ & { None} & { $h\to hh,V_L V_L\to V_L V_L h (4 V_L) $}  \\
${\cal O}_\psi$ & $y_\psi~\overline\psi_L H\psi_R(H^\dagger H)$ & {$h\to\psi\bar\psi$ 
} 
& { $V_LV_L\to t\bar t, V_L b\to t V_L V_L$} \\
${\cal O}_W$ & $\frac{i}{2}~g(H^\dagger\sigma^i\overleftrightarrow{D_\mu} H)(D_\nu W^{\mu\nu})^i$ & { $h\to VV^*, V^*\to hV$
}
 & { $q\bar q\to V_LV_L \, (h V_L)$} \\
${\cal O}_B$ & $\frac{i}{2}~g'(H^\dagger\overleftrightarrow{D_\mu} H)(\partial_\nu B^{\mu\nu})$ & { $h\to VV^*$ 
} & { $q\bar q\to V_LV_L  \, (h V_L)$}  \\
${\cal O}_{HW}$ & $ig(D^\mu H)^\dagger\sigma^i(D^\nu H)W_{\mu\nu}^i$ & $h\to \gamma Z$ 
& { $q\bar q\to V V$}  \\
${\cal O}_{HB}$ & $ig'(D^\mu H)^\dagger(D^\nu H)B_{\mu\nu}$ & $h\to \gamma Z$
& { $q\bar q\to VV$} \\
${\cal O}_{g}$ & $g_s^2H^\dagger H G^a_{\mu\nu}G^{a\mu\nu}$ & $h\to gg$ 
 & $pp\to V_LV_L,hh$ \\
${\cal O}_{\gamma}$ & $g'^2H^\dagger HB_{\mu\nu}B^{\mu\nu}$ & $h\to \gamma\gamma,\gamma Z,ZZ$ 
& $V_LV_L\to \gamma\gamma,\gamma Z,ZZ$ \\
\hline
${\cal O}_{2G}$ & $-\frac{1}{2}(D^\mu G_{\mu\nu})^a(D_\rho G^{\rho\nu})^a$ & None & { $pp\to jj$} \\
${\cal O}_{2W}$ & $-\frac{1}{2}(D^\mu W_{\mu\nu})^i(D_\rho W^{\rho\nu})^i$  & None & ${ q\bar q\to \psi\bar\psi}{ , VV}$ \\
${\cal O}_{2B}$ & $-\frac{1}{2}(\partial^\mu B_{\mu\nu})(\partial_\rho B^{\rho\nu})$ & None & ${ q\bar q\to \psi\bar\psi}{ ,VV}$ \\
${\cal O}_{3G}$ & $g_sf_{abc} G_\mu^{a\nu}G_\nu^{b\rho}G_{\rho}^{c\mu}$ & None & { $pp\to jj$} \\
${\cal O}_{3W}$ & $g \epsilon_{ijk} W_\mu^{i\nu}W_\nu^{j\rho}W_{\rho}^{k\mu}$ & None & { $q\bar q\to VV$}\\
\hline
\end{tabular}
}
\end{center}
\caption{\small Dimension-6 operators generated in universal theories \cite{Elias-Miro:2013eta,Wells:2015uba}, in the SILH basis of Ref.~\cite{Giudice:2007fh}. Here $y_\psi=\sqrt{2}m_\psi/v$ are the SM Yukawa couplings ($m_\psi$ is the fermion mass, and $v=246$ GeV the Higgs vev), $\lambda_h=m_h^2/(2v^2)$ the Higgs quartic (with $m_h$ its mass), $g_s$, $g$ and $g^\prime$ the $SU(3)$, $SU(2)$ and $U(1)$ gauge couplings. CP-odd versions of ${HW,HB,\gamma,g,3W,3G}$ can be obtained by replacing one field strength with the corresponding dual, $F^{\mu\nu}\to\frac{1}{2}\epsilon^{\mu\nu\alpha\beta}F_{\alpha\beta}$. The third column shows the dominant on-shell Higgs processes that the operator contributes to. The last column presents the cleanest processes enhanced at high energies. Here\label{tab1}}
\end{table}

\section{Weakly Coupled BSM}

\subsection[Minimal extensions of the Standard Model]{Minimal extensions of the Standard Model\footnote{\bf Based on a contribution by J. Criado, J. de Blas, M. Perez-Victoria, J. Santiago, following Ref.~\cite{deBlas:2017xtg}.}}\label{sec:minimalext}
\label{sec:operators}

In this note we use a bottom-up EFT approach to discuss new physics effects in Higgs physics 
in simple SM extensions with a few particles. We describe the part of the Standard Model Effective Field Theory (SMEFT) most important for Higgs physics and analyze the correlations found in each model
between the effects in Higgs physics and those in other types of observables.
The extra vector and fermion fields relevant for this study are listed in Table~\ref{tab:fields}, together with their SM quantum numbers. The effects of new scalars are discussed in section~\ref{sec:ESS_SMEFT}.

The operators in the Warsaw basis~\cite{Grzadkowski:2010es} that contain the Higgs field and can be generated
at the tree level are listed in the first column of table \ref{tab:bottom-up}. For simplicity we disregard
four-fermion interactions, even if they might influence Higgs physics indirectly.

The operators $\mathcal{O}_\phi$, $\mathcal{O}_{\phi \square}$ and
$(\mathcal{O}_{u \phi})_{33}$ are the ones that are currently less constrained by
experimental data.
The operator $\mathcal{O}_{\phi D}$ and  linear combinations of the operators of those of the type
$\mathcal{O}_{\phi \psi}$ have been constrained to be very small by electroweak
precision tests (EWPT), while $\mathcal{O}_{\phi ud}$ is also limited by low-energy data.
Experimental data from Higgs physics tell us that the Wilson coefficients of the interactions $(\mathcal{O}_{e\phi})_{33}$
and $(\mathcal{O}_{d\phi})_{33}$ should be well below 1 TeV$^{-2}$.

Using the dictionary in Ref.~\cite{deBlas:2017xtg}, we can readily identify which heavy fields 
can generate each operator at the tree level. They are listed in the second column of
table \ref{tab:bottom-up}, next to each corresponding operator. In the following, we consider a few simplified models that contain one or two of these fields. Our selection includes fields that appear frequently in more elaborate setups and illustrates typical features of the latter. Furthermore, all the operators in table~\ref{tab:bottom-up} are generated by this set of models.  We first discuss popular extensions of the SM with only one particle, which are severely constrained by EWPT. Then, we study minimal extensions with several particles that preserve custodial symmetry. In this case, the strongest constraints are evaded and strong effects in Higgs physics are allowed. In the explicit results below, it can be observed that many of the contributions to the Wilson coefficients have a definite sign. 

\begin{table}[t]
  \centering
  
  \begin{tabular}{ccccccc}
    \ctoprule
    \multirow{2}{*}{\bf Scalars} &
    \cc ${\cal S}$ &
    \cc $\varphi$ &
    \cc $\Xi$ &
    \cc $\Xi_1$ &
    \cc $\Theta_1$ &
    \cc $\Theta_3$ \\
    &
    $\left(1,1\right)_0$ &
    $\left(1,2\right)_{1/2}$ &
    $\left(1,3\right)_0$ &
    $\left(1,3\right)_1$ &
    $\left(1,4\right)_{1/2}$ &
    $\left(1,4\right)_{3/2}$ \\[1mm]
    \cbottomrule
  \end{tabular}

  \vspace{2mm}
  
  \begin{tabular}{cccccccc}
    \ctoprule
    \multirow{5}{*}{\bf Fermions} &
    \cc $N$ & \cc $E$ & \cc $\Delta_1$ & \cc $\Delta_3$ &
    \cc $\Sigma$ & \cc $\Sigma_1$ & \cc \\
    & $\left(1, 1\right)_0$ &
    $\left(1, 1\right)_{-1}$ &
    $\left(1, 2\right)_{-1/2}$ &
    $\left(1, 2\right)_{-3/2}$ &
    $\left(1, 3\right)_0$ &
    $\left(1, 3\right)_{-1}$ & \\[1mm]
    \cline{2-8} &
    \cc $U$ & \cc $D$ & \cc $Q_1$ & \cc $Q_5$ &
    \cc $Q_7$ & \cc $T_1$ & \cc $T_2$ \\
    & $\left(3, 1\right)_{2/3}$ &
    $\left(3, 1\right)_{-1/3}$ &
    $\left(3, 2\right)_{1/6}$ &
    $\left(3, 2\right)_{-5/6}$ &
    $\left(3, 2\right)_{7/6}$ &
    $\left(3, 3\right)_{-1/3}$ &
    $\left(3, 3\right)_{2/3}$ \\[1mm]
    \cbottomrule
  \end{tabular}

  \vspace{2mm}
  
  \begin{tabular}{cccccccc} 
    \ctoprule
    \multirow{2}{*}{\bf Vectors} &
    \cc ${\cal B}$ &
    \cc ${\cal B}_1$ &
    \cc ${\cal W}$ &
    \cc ${\cal W}_1$ \\
    & $\left(1,1\right)_0$ &
    $\left(1,1\right)_1$ &
    $\left(1,3\right)_0$ &
    $\left(1,3\right)_1$ \\[1mm]
    \cbottomrule
  \end{tabular}
  
  \caption{Representations of the fields with tree-level contributions
    to operators with the Higgs.}
  \label{tab:fields}
\end{table}

\vspace{1cm}

\begin{table}[h!]
  \begin{center}
    {\begin{tabular}{r@{\hspace{2pt}}lcl}
        \ctoprule
        & Name & Operator & Fields that generate it \\
        \cmrule
        $\star$ & $\mathcal{O}_{\phi}$ &
        $|H|^6$ &
        $\mathcal{S}$, $\varphi$, $\Xi$, $\Xi_1$, $\Theta_1$, $\Theta_3$,
        $\mathcal{B}_1$, $\mathcal{W}$ \\
        $\star$ & $\mathcal{O}_{\phi \square}$ &
        $|H|^2 \square |H|^2$ &
        $\mathcal{S}$, $\Xi$, $\Xi_1$,
        $\mathcal{B}$, $\mathcal{B}_1$, $\mathcal{W}$,
        $\mathcal{W}_1$ \\
        & $\mathcal{O}_{\phi D}$ &
        $|H^\dagger D_\mu H|^2$ &
        $\Xi$, $\Xi_1$,
        $\mathcal{B}$, $\mathcal{B}_1$, $\mathcal{W}$,
        $\mathcal{W}_1$ \\
        \cmrule
        $\bullet$ & $\mathcal{O}_{e \phi}$ &
        $|H|^2 \bar{l}_L H e_R$ &
        $\mathcal{S}$, $\varphi$, $\Xi$, $\Xi_1$,
        $E$, $\Delta_1$, $\Delta_3$, $\Sigma$, $\Sigma_1$,
        $\mathcal{B}$, $\mathcal{B}_1$, $\mathcal{W}$, $\mathcal{W}_1$
        \\
        $\bullet$ &  $\mathcal{O}_{d \phi}$ &
        $|H|^2 \bar{q}_L H d_R$ &
        $\mathcal{S}$, $\varphi$, $\Xi$, $\Xi_1$,
        $D$, $Q_1$, $Q_5$, $T_1$, $T_2$,
        $\mathcal{B}$, $\mathcal{B}_1$, $\mathcal{W}$, $\mathcal{W}_1$
        \\
        $\star$ &  $\mathcal{O}_{u \phi}$ &
        $|H|^2 \bar{q}_L \tilde{H} u_R$ &
        $\mathcal{S}$, $\varphi$, $\Xi$, $\Xi_1$,
        $U$, $Q_1$, $Q_7$, $T_1$, $T_2$
        $\mathcal{B}$, $\mathcal{B}_1$, $\mathcal{W}$, $\mathcal{W}_1$ \\
        \cmrule
        & $\mathcal{O}^{(1)}_{\phi l}$ &
        $(\bar{l}_L \gamma^\mu l_L)
        (H^\dagger i \overset{\leftrightarrow}{D}_\mu H)$ &
        $N$, $E$, $\Sigma$, $\Sigma_1$, $\mathcal{B}$ \\
        & $\mathcal{O}^{(3)}_{\phi l}$ &
        $(\bar{l}_L \gamma^\mu \sigma^a l_L)
        (H^\dagger i \overset{\leftrightarrow}{D}^a_\mu H)$ &
        $N$, $E$, $\Sigma$, $\Sigma_1$, $\mathcal{W}$ \\
        & $\mathcal{O}^{(1)}_{\phi q}$ &
        $(\bar{q}_L \gamma^\mu q_L)
        (H^\dagger i \overset{\leftrightarrow}{D}_\mu H)$ &
        $U$, $D$, $T_1$, $T_2$, $\mathcal{B}$ \\
        & $\mathcal{O}^{(3)}_{\phi q}$ &
        $(\bar{q}_L \gamma^\mu \sigma^a q_L)
        (H^\dagger i \overset{\leftrightarrow}{D}^a_\mu H)$ &
        $U$, $D$, $T_1$, $T_2$, $\mathcal{W}$ \\
        & $\mathcal{O}_{\phi e}$ &
        $(\bar{e}_R \gamma^\mu e_R)
        (H^\dagger i \overset{\leftrightarrow}{D}_\mu H)$ &
        $\Delta_1$, $\Delta_3$, $\mathcal{B}$ \\
        & $\mathcal{O}_{\phi u}$ &
        $(\bar{u}_R \gamma^\mu u_R)
        (H^\dagger i \overset{\leftrightarrow}{D}_\mu H)$ &
        $Q_1$, $Q_7$, $\mathcal{B}$ \\
        & $\mathcal{O}_{\phi d}$ &
        $(\bar{d}_R \gamma^\mu d_R)
        ( H^\dagger i \overset{\leftrightarrow}{D}_\mu H)$ &
        $Q_1$, $Q_5$, $\mathcal{B}$ \\
        & $\mathcal{O}_{\phi ud}$ &
        $(\bar{u}_R \gamma^\mu d_R) (H^\dagger i D_\mu \tilde{H})$ &
        $Q_1$, $\mathcal{B}_1$ \\
        \cbottomrule
      \end{tabular}}
  \end{center}
  \caption{Fields that generate each operator containing the Higgs at the tree
    level. Stars~$\star$ (bullets $\bullet$) indicate  operators that are less (more) constrained
    by experimental data, in comparison with other operators in the list (bullets  mark constraints that do not come from EWPT).}
  \label{tab:bottom-up}
\end{table}

\subsubsection{Models with one extra particle}
\label{sec:constrained}

\noindentFR{Quark singlet: $U \sim (3, 1)_{2/3}$}
\label{sec:qsinglet}
In this model, we have heavy vector-like quark with the same quantum numbers as the right-handed top:
\begin{align}
  \mathcal{L}_{\BSM} &= \mathcal{L}_{\SM} +
  i \bar{U} \slashed{D} U + M \bar{U} U
  - \left(
    \lambda_i \bar{U}_R \tilde{H}^\dagger q_{Li}
    + \text{h.c.}
  \right)\,,
\end{align}
where $q_{Li}$ are the SM left-handed quark doublets in the flavour eigenstate basis with diagonal SM up-quark Yukawa couplings.
To avoid flavour-changing neutral currents, we consider the case in which only one of the three $\lambda_i$ is non-vanishing.
The integration of the extra field out at the tree level, generates three
operators in the Warsaw basis, as presented in table
\ref{tab:qsinglet}. The first two operators contribute to gauge
couplings (also in association with one or two Higgs bosons) of the
SM quarks whereas the third one contributes to the up-type
Yukawa couplings (again plus one or two extra Higgs
bosons). A general analysis at the EFT level shows already that associated
$WH$ or $ZH$ production from the first two
operators in the case of first and second generation quarks is constrained by
EWPT. Other correlations appear once we consider this specific model.
Indeed, all three Wilson coefficients are controlled by a single parameter
$|\lambda_i|^2/M^2$. 
Top gauge couplings are not so severely constrained, 
thus allowing \textit{a priori} for a significant deviation of the top Yukawa
coupling. However, in this case EWPT still constrain the parameters
of the model through one-loop contributions. 

\begin{table}[t]
  \begin{center}
    \begin{tabular}{ccccc}
      \ctoprule
      $(C^{(1)}_{\phi q})_{ij}$ &
      $(C^{(3)}_{\phi q})_{ij}$ &
      $(C_{u \phi})_{ij}$ \\
      \cmrule
      $\frac{\lambda^*_i \lambda_j}{4 M^2}$ &
      $-\frac{\lambda^*_i \lambda_j}{4 M^2}$ &
      $2 y^{u*}_{jk} \frac{\lambda^*_i \lambda_k}{4 M^2}$ \\
      \cbottomrule
    \end{tabular}
  \end{center}
  \caption{Tree-level contributions to operators with the Higgs from the heavy
    top-like quark singlet. In the last coefficient $y^u$ denotes the SM Yukawa
    coupling for the up-type quarks.}
  \label{tab:qsinglet}
\end{table}

Current direct searches for pair production of this vector-like top
put a lower bound of about a TeV for its mass~\cite{Aaboud:2018pii}. Single production is also sensitive
to the Yukawa coupling of the new quarks~\cite{Aguilar-Saavedra:2013qpa}.

\noindentFR{Neutral vector singlet: $\mathcal{B} \sim (1, 1)_0$}
\label{sec:vsinglet}
In this model, a vector field $\mathcal{B}$ couples to the SM
fermions and the Higgs doublet:
\begin{align}
  \mathcal{L}_{\BSM} & =
  \mathcal{L}_{\SM}
  + \frac{1}{2} \left(
    (D_\mu \mathcal{B}_\nu) (D^\nu \mathcal{B}^\mu)
    - (D_\mu \mathcal{B}_\nu) (D^\mu \mathcal{B}^\nu)
    + M^2 \mathcal{B}_\mu \mathcal{B}^\mu
  \right) \nonumber \\ & \phantom{=}
  - \sum_{\psi} 
  (g^\psi)_{ij}
  \mathcal{B}^\mu \bar{\psi}_i \gamma_\mu \psi_j
  - \left[
    g^H \mathcal{B}^\mu H^\dagger iD_\mu H
    + \text{h.c.}
  \right].
\end{align}
The only component of this field is associated to a heavy neutral particle of spin 1, i.e. a $Z^\prime$ boson. 
We work in the field basis with diagonal and canonical kinetic terms and assume that the coupling constant $g^H$ to the Higgs is real. We describe massive spin-1 fields as Proca vector fields; the origin of the mass term is not relevant for our purposes here as long as any other possible particle in a complete model (such as a symmetry-breaking scalar) is sufficiently heavier.
Tree-level matching to the SMEFT gives four-fermion operators plus the
contributions that appear in table \ref{tab:vsinglet}. From the EWPT
constraints on $\mathcal{O}_{\phi D}$, it follows that $g^H$
should be small. This affects every operator with the Higgs generated by
$\mathcal{B}$, which will be suppressed by $g^H$ or
$(g^H)^2$.

\begin{table}[h]
  \begin{center}
    \begin{tabular}{ccccc}
      \ctoprule
      $C_{\phi D}$ &
      $C_{\phi \square}$ &
      $(C^{[(1)]}_{\phi \psi})_{ij}$ \\
      \cmrule
      $- \frac{2 (g^H)^2}{M^2}$ &
      $- \frac{(g^H)^2}{2 M^2}$ &
      $- \frac{
        (g^\psi)_{ij} (g^H)}{M^2}$ \\
      \cbottomrule
    \end{tabular}
    \caption{Tree-level contributions to operators with the Higgs from the
      neutral vector singlet.}
    \label{tab:vsinglet}
  \end{center}
\end{table}

Searches for single production of neutral vectors decaying to dileptons,
dibosons or dijets exclude additional regions in the parameter space of this model~\cite{Sirunyan:2018exx,Aad:2019fac,Sirunyan:2018xlo}.

\subsubsection{Custodial models}
\label{sec:unconstrained}

\noindentFR{Quark bidoublet: $Q_1 \sim (3, 2)_{1/6}$ and $Q_7 \sim (3, 2)_{7/6}$}
\label{sec:bidoublet}
We introduce two quark doublets,
\begin{align}
  Q_7 =
  \left( \begin{array}{c} X \\ T \end{array} \right),
  \qquad
  Q_1 =
  \left( \begin{array}{c} T' \\ B \end{array} \right),
\end{align}
with the same mass and coupling to the top quark:
\begin{align}
  \mathcal{L}_{\BSM} & =
  \mathcal{L}_{\SM} +
  i \bar{Q}_7 \slashed{D} Q_7
  + i \bar{Q}_1 \slashed{D} Q_1
  \nonumber \\ & \phantom{=}
  + M \left(
    \bar{Q}_7 Q_7 + \bar{Q}_1 Q_1
  \right)
  \nonumber \\ &
  \phantom{=} - \left[
    \lambda \left(
      \bar{Q}_{7L} H t_R
      + \bar{Q}_{1L} \tilde{H} t_R
    \right)
    + \text{h.c.}
  \right].
  \label{eq:LQ1Q7}
\end{align}

This Lagrangian respects custodial symmetry, with the pair of quarks
transforming under a $(2, 2)_{2/3}$ representation of $SU(2)_L \times SU(2)_R \times U(1)_X$,
where the hypercharge is $Y = T^R_3 + X$. The contributions to the
$\mathcal{O}_{\phi \psi}$ operators from both doublets cancel each other.
Only the operator $(\mathcal{O}_{u \phi})_{33}$ is generated by a tree-level integration, with a positive Wilson coefficient. 
The explicit value of $(C_{u \phi})_{33}$ in this SM extension is given in table~\ref{tab:Qbidoublet}.

\begin{table}[h]
  \begin{center}
    \begin{tabular}{ccc}
      \ctoprule
      &$(C_{u \phi})_{33}$& \\
      \cmrule
      &$\frac{|\lambda|^2}{M^2}$& \\
      \cbottomrule
    \end{tabular}
    \caption{Tree-level contributions to operators with the Higgs from the
      quark doublets $Q_7$ and $Q_1$, with the interactions in Eq.~(\ref{eq:LQ1Q7}).}
    \label{tab:Qbidoublet}
  \end{center}
\end{table}

Therefore, this is a model which can give large negative contributions to the top
Yukawa coupling without producing any other effects at the tree level. 
Note that one-loop constraints are under control for this particular model: 
contributions to the $T$ parameter are protected by custodial symmetry, 
bounds from the $S$ parameter 
are mild, and the contributions of the new quarks to $\mathcal{O}_g$ and $\mathcal{O}_\gamma$ compensate the reduction in the top Yukawa coupling~\cite{Aguilar-Saavedra:2013qpa}.

The mass of the extra quarks is bounded from below by direct pair production limits, similarly to the case of the singlet (section \ref{sec:qsinglet})~\cite{Aaboud:2018pii}.

\noindentFR{Neutral vector triplet: $\mathcal{W} \sim (1, 3)_0$}
\label{sec:vtriplet}
The neutral vector triplet contains a $Z'$ and a $W'$. It couples to the SM
doublets:
\begin{align}
  \mathcal{L}_{\BSM} & = \mathcal{L}_{\SM}
  + \frac{1}{2} \left(
    (D_\mu \mathcal{W}^a_\nu) (D^\nu \mathcal{W}^{\mu a})
    - (D_\mu \mathcal{W}^a_\nu) (D^\mu \mathcal{W}^{\nu a})
  \right)
  \nonumber \\ & \phantom{=}
  + \frac{1}{2} M^2 \mathcal{W}^a_\mu \mathcal{W}^{\mu a}
  - \frac{1}{2} (g^l)_{ij}
  \mathcal{W}^{\mu a} \bar{l}_{Li} \sigma^a \gamma_\mu l_{Lj}
  \nonumber \\ & \phantom{=}
  - \frac{1}{2} (g^q)_{ij}
  \mathcal{W}^{\mu a} \bar{q}_{Li} \sigma^a \gamma_\mu q_{Lj}
  - \left[
    \frac{1}{2} (g^H)
    \mathcal{W}^{\mu a} H^\dagger \sigma^a iD_\mu H
    + \text{h.c.}
  \right].
\end{align}

We assume that $g^H$ is real. Table \ref{tab:vtriplet}
summarizes the tree-level contributions of $\mathcal{W}$ to operators with the
Higgs. Unlike the case of the vector singlet ${\cal B}$, the coupling $g^{H}$ is allowed 
to be large in this case, because the
contribution to the $T$ parameter is zero. Therefore, in this model there can
be large contributions controlled by $g^H$ to double Higgs
production (from $\mathcal{O}_\phi$), the Higgs kinetic term (from
$\mathcal{O}_{\phi \square}$) and the SM Yukawa couplings (from
$\mathcal{O}_{\psi \phi}$).

\begin{table}[h]
  \begin{center}
    \begin{tabular}{cccccccc}
      \ctoprule
      $C_{\phi}$ &
      $C_{\phi \square}$ &
      $(C_{\psi \phi})_{ij}$ &
      $(C^{(3)}_{\phi \psi})_{ij}$ \\
      \cmrule
      $- \frac{\lambda_H (g^H)^2}{M^2}$ &
      $- \frac{3 (g^H)^2}{8 M^2}$ &
      $- \frac{y^{\psi*}_{ji} (g^H)^2}{4 M^2}$ &
      $- \frac{
        (g^\psi)_{ij} (g^H)}{
        4 M^2}$ \\
      \cbottomrule
    \end{tabular}
  \end{center}
  \caption{Tree-level contributions to operators with the Higgs from the
    neutral vector triplet. In the first and third coefficients, $\lambda_H$ and $y^\psi$ denote the SM 
    Higgs doublet quartic interaction and fermion Yukawa
    couplings, respectively.
    }
  \label{tab:vtriplet}
\end{table}

As for the singlet (section \ref{sec:vsinglet}), direct searches for single
production of $Z'$ and $W'$ apply here~\cite{Aaboud:2018bun}.

\noindentFR{Pair of vector singlets: $\mathcal{B} \sim (1, 1)_0$ and
  $\mathcal{B}_1 \sim (1, 1)_1$}
\label{sec:vpair}
In this model we have two singlet vector fields with different hypercharges. The $\mathcal{B}$ field contains a $Z'$, as already indicated, and the $\mathcal{B}_1$, a $W'$ with right-handed couplings. We assign to them the same mass and related couplings to the Higgs, in such a way that they combine to form a $(1, 3)_0$ representation of $SU(2)_L \times SU(2)_R \times U(1)_X$ and custodial symmetry is preserved:
\begin{align}
  \mathcal{L}_{\BSM} & =
  \mathcal{L}_{\SM}
  + \mathcal{L}^{\mathcal{B}}_{\text{kin}}
  + \mathcal{L}^{\mathcal{B}_1}_{\text{kin}}
  \nonumber \\ & \phantom{=}
  + M^2 \left(
    \frac{1}{2} \mathcal{B}_\mu \mathcal{B}^\mu
    + \mathcal{B}^\dagger_{1\mu} \mathcal{B}^{\mu}_1
  \right)
  \nonumber \\ &
  - \left[
    g^H \left(
      \mathcal{B}^\mu H^\dagger iD_\mu H
      + \sqrt{2} \mathcal{B}^{\mu\dagger}_1 i D_\mu H^T i\sigma_2 H
    \right)
    + \text{h.c.}
  \right].
\end{align}
For simplicity, we have not included in the previous equation the fermionic couplings to the heavy vectors. 
Their indirect effects are independent from the ones discussed in this section.

The
results of integrating the extra fields out at the tree level appear in table
\ref{tab:vpair}. The contribution to $\mathcal{O}_{\phi D}$ from
$\mathcal{B}_1$ cancels the one from $\mathcal{B}$. This means that the limits on $g^H$ are milder than in the case with $\mathcal{B}$ alone (section
\ref{sec:vsinglet}). In fact, this coupling is not constrained by EWPT when the fermion couplings vanish. Therefore, large effects
in $\mathcal{O}_\phi$, $\mathcal{O}_{\phi \square}$ and $\mathcal{O}_{\psi \phi}$ are allowed in this model.

Direct searches at the LHC are sensitive to the Higgs couplings of the new vectors via vector boson production
of the new fields and decays into dibosons~\cite{Chatrchyan:2012rva}. Other channels are typically more restrictive when the couplings to fermions are not small.

\begin{table}[h]
    \begin{center}
    \begin{tabular}{ccc}
      \ctoprule
      $C_{\phi}$ &
      $C_{\phi \square}$ &
      $(C_{\psi \phi})_{ij}$ \\
      \cmrule
      $- \frac{4 \lambda_H (g^H)^2}{M^2}$ &
      $- \frac{3 (g^H)^2}{2 M^2}$ &
      $- \frac{y^{\psi*}_{ji} (g^H)^2}{M^2}$ \\
      \cbottomrule
    \end{tabular}
  \end{center}
  \caption{Tree-level contributions to operators with the Higgs from the
    pair of vector singlets. In the first and third coefficients, $\lambda_H$ and $y^\psi$ denote the SM 
    Higgs doublet quartic interaction and fermion Yukawa
    couplings, respectively.}
  \label{tab:vpair}
\end{table}

\subsection[Extended Scalar Sectors and the SMEFT]{Extended Scalar Sectors and the SMEFT\footnote{\bf Based on a contribution by  S.~Dawson and C.~Murphy, following Ref.~\cite{Dawson:2017vgm}.}}
\label{sec:ESS_SMEFT}

In this section we focus on extensions of the SM involving heavy scalars and address the impact of these models for Higgs phenomenology, keeping into account stringent constraints from electroweak precision tests via the $\rho$-parameter ($T$-parameter). 

The dimension-6 operators of interest for this work are 
\begin{align}
\label{eq:Left}
\mathcal{L}^{(6)}_H= \frac{\hat c_H}{v^2} \mathcal{O}_H + \frac{\hat c_T}{ v^2}  \mathcal{O}_T  - \frac{\hat c_6 }{v^2}  \mathcal{O}_6  
+ \frac{\left(H^{\dagger} H\right)}{v^2} \left[\hat c_b  \mathcal{O}_b + \hat c_t  \mathcal{O}_t + \hat c_{\tau}  \mathcal{O}_\tau+ \text{h.c.}\right] , 
\end{align}
where the operators in the SILH basis have been defined in table \ref{tab1} ($\mathcal{O}_b$, $\mathcal{O}_t$ and $\mathcal{O}_\tau$ correspond to $\mathcal{O}_\psi$, with $\psi=b,t,\tau$ respectively), and we have normalized them as $\hat c_i=c_i v^2/\Lambda^2$ compared to \eq{eq:L_SMEFT}.

These dimension-6 operators modify the Higgs boson self-interactions
\begin{align}
\label{eq:LeftCan}
\mathcal{L}_h &= \frac{1}{2} \left(\partial_{\mu} h\right)^2 - \frac{1}{2} m_h^2 h^2  - \frac{m_h^2}{2 v} \left(1 + \hat c_6 - \frac{3}{2} \hat c_H\right) h^3 \\
&- \frac{m_h^2}{8 v^2} \left(1 + 6 \hat c_6 - \frac{25}{3} \hat c_H\right) h^4 - \frac{m_h^2}{48 v^4} \left(3 \hat c_6 - 4 \hat c_H\right) h^5 \left(6 v + h\right) , \nn 
\end{align}
and contribute to the $\rho$ parameter~\cite{Ross:1975fq}
\begin{equation}
\label{eq:rho6}
\rho = 1 + \hat c_T .
\end{equation}
They also modify  the Yukawa couplings of the Higgs boson
\begin{equation}
\mathcal{L}_{y_t} = - m_t \bar{t} t \left[1 + \left(1 - \frac{\hat c_H + 2 \hat c_t}{2}\right) \frac{h}{v} - \frac{\hat c_H + 3 \hat c_t}{2} \left(\frac{h^2}{v^2} + \frac{h^3}{3 v^3}\right)\right] ,
\end{equation}
and similarly for the other SM fermions.


We consider a number of extensions of the SM where a single new color singlet, spin-zero multiplet, $\phi$,  is added to the SM and require that there is a renormalizable interaction with the SM $H$ doublet that is linear in $\phi$. 
There is a sizable literature on integrating out heavy scalars and studying their SM Effective Field Theory (SMEFT) contributions, see for instance Ref.s~\cite{Khandker:2012zu, Henning:2014wua, deBlas:2014mba, Gorbahn:2015gxa, Chiang:2015ura, Brehmer:2015rna, Egana-Ugrinovic:2015vgy, delAguila:2016zcb, Buchalla:2016bse, Belusca-Maito:2016dqe, Jiang:2016czg, Corbett:2017ieo, deBlas:2017xtg}. 

The potential can schematically be written as 
\begin{equation}
V\left(H, \phi\right) = V_{SM}\left(H\right) + V_{Z_2}\left(H, \phi\right) + V_{\bcancel{Z_2}}\left(H, \phi\right) ,
\end{equation}
where $\phi$ is the new scalar, and $V_{SM}$ is the SM potential.
For a real valued $\phi$, the $Z_2$ preserving potential has the following form
\begin{equation}
V_{Z_2}\left(H, \phi\right) = \frac{1}{2} M^2 \phi^a \phi^a + \lambda_{\alpha} H^{\dagger} H \phi^a \phi^a + \lambda_{\beta} \left(\phi^a \phi^a\right)^2 ,
\label{eq:z2}
\end{equation}
where $a$ are the $SU(2)_L$ indices,  and for a complex valued $\phi$ there may be multiple $\alpha$ and/or $\beta$-type interactions. 
Additionally, when $\phi$ is complex, there is no factor of one-half in front of the mass term, and $\phi^a \phi^a$ should be replaced with $\phi^{a \dagger} \phi^a$. Depending on the $SU(2)_L$  representation of $\phi$, the $Z_2$ violating potential contains one of the following interactions
\begin{equation}
 V_{\bcancel{Z_2}} \sim m_1 H^2 \phi \quad \text{or} \quad V_{\bcancel{Z_2}} \sim \lambda_1 H^3 \phi .
 \label{eq:noz2}
 \end{equation}
If $\phi$ is a singlet there can also be a tadpole term and a cubic self-interaction, both of which violate the $Z_2$ symmetry. 
From Ref.~\cite{Dawson:2017vgm} we see that taking $\lambda_1$ or $m_1 \to 0$  (Eq. \ref{eq:noz2}) 
while holding the other parameters fixed, or sending $M \to \infty$ (Eq. \ref{eq:z2})  also while keeping the other parameters fixed, causes the new scalar multiplet to decouple.
These are the analogs of the alignment without decoupling limit, and  the decoupling limit of the 2HDM, respectively~\cite{Gunion:2002zf, Carena:2013ooa}. 
 
We define the angle $\alpha$ to characterize the mixing between the neutral, $CP$-even components of $H$ and $\phi$
\begin{equation}
\label{eq:singmix}
\begin{pmatrix}
h \\
{\cal{H}}
\end{pmatrix} = 
\begin{pmatrix}
\cos\alpha & - \sin\alpha \\
\sin\alpha & \cos\alpha
\end{pmatrix}
\begin{pmatrix}
h^\prime \\
\varphi
\end{pmatrix} ,
\end{equation}
where $\text{Re}(H) = {v_h + h^\prime\over\sqrt{2}}$ , $\text{Re}(\phi^0) = v_{\phi} + \varphi$,  $v_h$ and $v_{\phi}$ are the vevs of $H$ and $\phi$, respectively, and $h$ and ${\cal{H}}$ are the physical Higgs fields.  
In the 2HDM there is an ambiguity as both scalar fields have the same quantum numbers.
Our results straightforwardly apply to the so-called Higgs basis of the 2HDM, where our $H$ and $\phi$ fields respectively correspond to the $H_1$ and $H_2$ fields of \textit{e.g.} Ref.~\cite{Davidson:2005cw}.
The relevant angle in the 2HDM is $\beta - \alpha$, where $\alpha$ has the same interpretation as in Eq.~\eqref{eq:singmix}, and $\beta$ will be defined below.
In all of the models we consider, a non-zero value of $\alpha$ leads to a universal modification of the Higgs couplings to SM particles (excluding the Higgs self-couplings). 

With the above definitions of the vevs of $H$ and $\phi$, the electroweak (EW) vev is given by
\begin{equation}
v^2 = v_h^2 + 2 \left[t(\phi)\left(t(\phi) + 1\right) - t_3(\phi)^2\right] v_{\phi}^2 ,
\end{equation}
where $t(i)$, $t_3(i)$, and $v_i$ are the representations under $SU(2)_L$ of the $i$th multiplet, the neutral component of the $i$th multiplet, and the vev of the $i$th multiplet, respectively.
When $\phi$ is a singlet $v_h = v$, and we define $\tan\beta_s = v_h / v_{\phi}$.
For higher $SU(2)_L$ representations, we define the mixing angle between the two vevs as
\begin{equation}
\label{eq:betadef}
v_h = v \cos\beta, \quad v_{\phi} = v \sin\beta / \sqrt{2 \left[t(\phi)\left(t(\phi) + 1\right) - t_3(\phi)^2\right]} .
\end{equation}
In models that generate $\hat c_T$ the mixing angle $\beta$ is constrained by measurements of the $\rho$ parameter, see Table~\ref{tab:rho}.
\begin{table}
\centering
 \begin{tabular}{| c | c | c |}
 \hline 
\textbf{Model} & $\mathbf{\rho}$ & \textbf{$3\sigma$ upper limit on $\beta$}  \\ \hline 
Singlet & 1 & none \\ \hline
2HDM & 1 & none \\ \hline
Real Triplet & $\sec^2\beta$ & 0.030  \\ \hline
Complex Triplet & $2\left(3 - \cos2\beta\right)^{-1}$ & 0.014   \\ \hline
Quartet: $Y = \tfrac{1}{2}$ & $7\left(4 + 3 \cos2\beta\right)^{-1}$ & 0.033  \\ \hline
Quartet: $Y = \tfrac{3}{2}$ & $\left(2 - \cos2\beta\right)^{-1}$ & 0.010  \\ \hline
 \end{tabular}
  \caption{The tree level contribution to $\rho$ in a given model, and the corresponding $3\sigma$ upper limit on the mixing angle $\beta$.
  Reproduced from Ref.~\cite{Dawson:2017vgm}.}
  \label{tab:rho}
\end{table}

Given these interactions, standard methods exist to determine which operators in the SMEFT are generated at tree level in a given model~\cite{Khandker:2012zu, Henning:2014wua}.
These results are compiled in Table~\ref{tab:dim6ops_note}.
The singlet model coefficients are expressed in terms of the mixing angle between $h$ and $\mathcal{H}$.
In the singlet and triplet models the constant of the proportionality that relates $\hat c_6$ to the other Wilson coefficients is a free parameter that may be sizable when the UV theory is strongly interacting.
The 2HDM coefficients depend on both mixing angles $\alpha$ and $\beta$. 
The cutoff scale appearing in the expressions for $\hat c_6$ in the 2HDM and quartet models is a common mass for the heavy Higgses.
The triplets and quartets contribute to the $\rho$ parameter at tree level, and we express all the coefficients in terms of a given model's contribution to $\hat c_T$.

\begin{table}
\centering
\small
 \begin{tabular}{| c | c | c | c | c | c |}
 \hline 
\textbf{Model} & $\mathbf{\hat c_H}$ & $\mathbf{\hat c_6 \lambda_{SM}}$ & $\mathbf{\hat c_T}$ & $\mathbf{\hat c_t}$ & $\mathbf{\hat c_b} = \mathbf{\hat c_{\tau}}$ \\ \hline 
Real Singlet: exp. $\bcancel{Z_2}$ & $\tan^2\alpha$ & $\tan^2\alpha \left(\lambda_{\alpha} \!\!-\!\! \frac{m_2}{v} \tan\alpha\right)$ & 0 & 0 & 0    \\ \hline
Real Singlet: spont. $\bcancel{Z_2}$ & $\tan^2\alpha$ & 0 & 0 & 0  & 0  \\ \hline
2HDM: Type I & 0 & $- c_{\beta-\alpha}^2 \frac{\Lambda^2}{v^2}$ & 0 & $- c_{\beta-\alpha} \cot\left(\beta\right)$ & $- c_{\beta-\alpha}\cot\left(\beta\right)$    \\ \hline
2HDM: Type II & 0 & $- c_{\beta-\alpha}^2 \frac{\Lambda^2}{v^2}$ & 0 & $- c_{\beta-\alpha} \cot\left(\beta\right)$ & $c_{\beta-\alpha}\tan\left(\beta\right)$   \\ \hline
Real Triplet & $- 2 \hat c_T$ & $\hat c_T \lambda_{\alpha}$ & \checkmark & $\hat c_T$ & $\hat c_T$  \\ \hline
Complex Triplet & $\hat c_T$ & $- \hat c_T \left(\lambda_{\alpha 1} - \frac{\lambda_{\alpha 2}}{2}\right)$ & \checkmark & $- \hat c_T$ & $- \hat c_T$  \\ \hline
Quartet: $Y = \tfrac{1}{2}$ & 0 & $- 2 \hat c_T \frac{\Lambda^2}{v^2}$ & \checkmark & 0 & 0  \\ \hline
Quartet: $Y = \tfrac{3}{2}$ & 0 & $\frac{2}{3} \hat c_T \frac{\Lambda^2}{v^2}$ & \checkmark & 0 & 0  \\ \hline
 \end{tabular}
  \caption{The dimension-6 operators from Eq.~\eqref{eq:Left} that are generated at tree level in the models under consideration in the decoupling limit. Here $c_{\beta-\alpha}\equiv  \cos\left(\beta - \alpha\right)$. 
Adapted from Ref.~\cite{Dawson:2017vgm}.}
  \label{tab:dim6ops_note}
\end{table}

\subsection[One-loop BSM]{One-loop BSM\footnote{\bf Based on a contribution by Brian Henning, following Ref.~\cite{Henning:2014wua}.}}
\label{sec:loop}
We continue our discussion of perturbative UV models, except here we focus on effects which arise at one-loop level. More precisely, we consider cases in which tree-level effects are absent. Tree-level effects come from operators which are linear in a heavy field \(\Phi\), \(\mathcal{L}_{\text{UV}} \supset \Phi \mathcal{O}[\phi_{\text{SM}}]\). The absence of such terms is usually a consequence of symmetry, be it because of the SM gauge symmetry or some global symmetry such as a discrete \(Z_2\) acting on \(\Phi\).\footnote{An example of the first type are scalar tops in SUSY; examples of the latter type include \(R\)-parity in SUSY, KK-parity in extra-dimensional models, T-parity in little Higgs models, and the inert 2HDM.} A full classification of tree- and loop-level effects can be found in~\cite{Arzt:1994gp}.

One-loop effects originate out of operators which are \textit{quadratic} in the heavy fields, \(\Phi_i\Phi_j\mathcal{O}[\phi_{\text{SM}}]\).\footnote{An exception can arise if there are also linear terms (\textit{i.e.} tree-level effects). In such a case, higher order self-interactions of the \(\Phi_i\) can contribute at one-loop; diagramatically, these are \textit{not} one-particle irreducible Feynman diagrams. See, \textit{e.g.},~\cite{Henning:2014wua,Henning:2016lyp}.} Our assumption that the UV models under consideration are weakly coupled at the scale of the heavy particles means that we can focus on operators of mass dimension \(\le 4\) in the UV Lagrangian. By dimensional analysis, this further implies that we need only consider BSM states which are scalars (\(S\), with canonical mass dimension \([S] = 1\)), fermions (\(F\), with  \([F] = \frac{3}{2}\)), or vectors (\(V\), with \([V] = 1\)).

Our interest lies in how the BSM states modify the Higgs sector, \textit{i.e.} couplings between the Higgs doublet \(H\) and the \(\Phi_i = S_i\), \(F_i\), or \(V_i\). It is very important to recognize that this generically modifies the EW gauge sector as well since, in order to write down gauge-invariant interactions, the \(\Phi_i\) will typically carry EW quantum numbers. With these considerations, it is easy to schematically write down all possible UV interactions that (1) modify the Higgs and/or EW sector, (2) arise at one-loop (quadratic in heavy fields), and (3) are relevant or marginal interactions (mass dimension \(\le 4\)):
\begin{equation}\label{eq:list_schematic}
  \begin{array}{ccccc}
    H S^2 & ~~~ & \bar{F} \slashed{D} F & ~~~ & H^2 V^2 \\
    H^2 S^2 & & \xcancel{H F^2} & & (DV)^2 \\
    (DS)^2 & & & & 
  \end{array}.
\end{equation}
Here \(D\) denotes the covariant derivative of the SM, so that operators involving \(D\) indicate the potential coupling of BSM states to SM gauge bosons. We have crossed out the Yukawa-like terms \(H F^2\) since the SM gauge symmetry does not allow these interactions if we require the fermions to have a large mass not originating from EW symmetry breaking (\textit{i.e.} when we require the fermions to be vector like, with a mass \(M_F \gg v\)).

Below we will give specific examples of motivated BSM models containing interactions of the type in eq.~\eqref{eq:list_schematic}, detailing how these manifest in the SM EFT. First, however, we give some general comments on matching the UV models to the SM EFT, emphasizing estimates of the size the EFT Wilson coefficients and general patterns of deviation one can look for to get insights into the new physics.

\subsubsection{Integrating out, power counting, and general patterns}
The imprints that BSM states from a given  UV model leave on EW physics can be quantified by integrating out the BSM states and matching onto the SM EFT. That is, in the SM EFT \eq{eq:L_SMEFT}, we want to find the resultant Wilson coefficients \(c_i\) when we integrate out the heavy BSM states. This process of ``matching'' onto the EFT is done by requiring that the UV model and the EFT give the same physics at the scale \(\mu = \Lambda\) where the heavy states are integrated out (in technical terms, this is achieved by equating the one-light-particle-irreducible effective actions \(\Gamma_{L,\text{UV}}( \Lambda) = \Gamma_{L,\text{EFT}}(\Lambda)\), \textit{e.g.}~\cite{Georgi:1991ch}). The assumption that the BSM physics is weakly coupled means that the \(c_i\) in \eq{eq:L_SMEFT} have a well-defined expansion in powers of \(\hbar\),
\begin{equation}\label{eq:wilson_hbar}
  c_i = c_i^{\text{tree}} + c_i^{\text{1-loop}} + \dots,
\end{equation}
where \(c_i^{\text{tree}}\) arises at tree-level and have been discussed so far, \(c_i^{\text{1-loop}}\) at one-loop, \textit{etc}.

In the SM EFT, the most relevant operators giving deviations to the Higgs and EW gauge sector begin at dimension six. For weakly coupled BSM physics, a rough estimate of the size of the Wilson coefficients is straightforward: the power-counting is essentially just the loop counting of eq.~\eqref{eq:wilson_hbar}. So, if a heavy BSM state of mass \(\Lambda\) generates some dimension-six operator we expect
\begin{subequations}
  \begin{align}
    c_i^{\text{tree}} &\approx \frac{g_{\text{UV}}^{\#}}{\Lambda^2}, \\
    c_i^{\text{1-loop}} &\approx \frac{1}{16\pi^2}\frac{g_{\text{UV}}^{\#}}{\Lambda^2},
  \end{align}
\end{subequations}
where \(g_{\text{UV}}\) generically denotes products of coupling constants in the UV Lagrangian (in certain models, UV couplings coincide with SM couplings, $g_{\text{UV}}=g_{\text{SM}}$).\footnote{To compute their effects on a weak-scale observable, the Wilson coefficients need to be RG evolved down to the scale \(\mu \approx v\). The RG running introduces another loop factor times a log, \(\frac{1}{(4\pi)^2}\log\left( \frac{\Lambda}{\mu}\right)\), and involves powers of the SM coupling constants. The references~\cite{Grojean:2013kd,Jenkins:2013zja,Elias-Miro:2013gya,Alonso:2013hga,Elias-Miro:2013mua} compute the anomalous dimension matrix necessary to perform this RG running. Notice also that, if \(c_i^{\text{tree}} \ne 0\), then it is possible for \(c_i^{\text{1-loop}}\) to also have contributions proportional to powers of UV and SM coupling constants. Such contributions arise from diagrams involving both heavy and light particles propagating in the loop, see~\cite{delAguila:2016zcb,Henning:2016lyp,Ellis:2016enq,Fuentes-Martin:2016uol,Zhang:2016pja,Ellis:2017jns} for further discussion and examples.}

For the present discussion we consider BSM theories whose leading order effect begins at one-loop, \textit{i.e.} for which all \(c_i^{\text{tree}} =0\) (see~\cite{delAguila:2016zcb,Henning:2016lyp,Ellis:2016enq,Fuentes-Martin:2016uol} for the more general case). We recall that the one-loop effects arise out of interactions in the BSM model which are quadratic in the heavy particles, eq.~\eqref{eq:list_schematic}. Functional methods provide an efficient way to match onto the low-energy EFT, see \textit{e.g.}~\cite{Henning:2014wua} for an introduction. Here, one integrates out the heavy fields from the UV action \(S_{\text{UV}}\). Integrating out the heavy field(s) \(\Phi\) at one-loop leads to the effective action~\cite{Henning:2014wua}
\begin{equation}
  \Delta S_{\text{eff,1-loop}} = i c_s \text{Tr} \log \left( -\frac{\delta^2 S_{\text{UV}}}{\delta \Phi^2} \right)= i c_s \text{Tr}\log \Big[D^2 + M_{\Phi}^2 + U_s[\phi_{\text{SM}}] \Big].
  \label{eq:general_functional}
\end{equation}
In this equation \(D\) is the covariant derivative, \(M_{\Phi}\) is the mass (matrix) of the heavy field(s), \(U_s[\phi_{\text{SM}}]\) comes from the quadratic interactions of \(\Phi\) with the SM fields \(\phi_{\text{SM}}\), and \(c_s\) is a constant depending on the spin of \(\Phi_i\) (respectively, \(c_s = \frac{1}{2}, 1, -\frac{1}{2}, \frac{1}{2}\) for real scalars, complex scalars, Dirac fermions, and gauge bosons).

The functional trace is subsequently evaluated in a way that allows for an inverse mass expansion---that is, in powers of \(1/\Lambda\)---thereby producing a series of higher-dimension operators from which one reads off the corresponding Wilson coefficients. One of the nice features about functional methods is that they can be evaluated in a gauge-covariant manner~\cite{Gaillard:1985uh,Cheyette:1987qz} (\textit{i.e.} one can keep the covariant derivative \(D_{\mu} = \partial_{\mu} + i A_{\mu}\) intact throughout the calculation); see~\cite{Henning:2014wua,Drozd:2015rsp,Zhang:2016pja} or the summary in~\cite{Brivio:2017vri}.

\subsubsection{Examples}\label{sec:oneloop_examples}

\paragraph{Stops: $HS^2,\ H^2S^2,\ (DS)^2$} \textcolor{white}{.}

Here we consider the resulting Wilson coefficients from integrating out heavy stops from the MSSM~\cite{Henning:2014gca}. Arranging the stops into the multiplet \(\Phi = \begin{pmatrix} \tilde{Q}_3 \\ \tilde{t}_R \end{pmatrix}\), the relevant quadratic piece of the MSSM Lagrangian is
\begin{equation}
\mathcal{L}_{\text{MSSM}} \supset \Phi^\dagger \left( -D^2-M^2-U \right) \Phi ,
\end{equation}
where the mass and \(U\) matrices entering eq.~\eqref{eq:general_functional} are
\begin{equation}
{M^2} = \left( {\begin{array}{*{20}{c}}
{m_{{{\tilde Q}_3}}^2}&0\\
0&{m_{{{\tilde t}_R}}^2}
\end{array}} \right) ,
\end{equation}
and
\begin{equation}
U = \begin{pmatrix} \big(y_t^2s_{\beta}^2 + \frac{1}{2}g^2c_{\beta}^2\big)\tilde{H}\tilde{H}^{\dag} + \frac{1}{2} g^2 s_{\beta}^2HH^{\dag} - \frac{1}{2}\big(g'^2 Y_Q c_{2\beta} + \frac{1}{2}g^2\big) \left|H\right|^2 & y_t s_{\beta} X_t \tilde{H} \\
  y_t s_{\beta} X_t \tilde{H}^{\dag} & \big( y_t^2s_{\beta}^2 - \frac{1}{2} g'^2 Y_{t_r} c_{2\beta}\big) \left|H\right|^2 \end{pmatrix}.
\end{equation}
In the above, \(\tilde{H} \equiv i \sigma_2 H^*\); \(g\), \(g'\), and \(y_t\) are SM couplings (see definitions in the caption of Table~\ref{tab1}, with $\psi=t$); and \(\tan \beta \equiv \langle H_u \rangle / \langle H_d \rangle \) is the usual ratio of vevs in the MSSM.

Assuming degenerate stop masses for simplicity \(m_{\tilde{Q}_3} = m_{\tilde{t}_R} \equiv m_{\tilde{t}}\), then the resulting Wilson coefficients are summarized in table~\ref{tbl:MSSM}.\footnote{The results for non-degenerate stop masses can be found in~\cite{Drozd:2015kva}.}
In this table, as well as the rest of this subsection, the normalization for \(\mathcal{O}_6\) and \(\mathcal{O}_{2G,2W,2B}\) differs slightly from table~\ref{tab1}: here we do not include the multiplicative SM couplings in the definition of the operator (\textit{e.g.} here \(\mathcal{O}_6 = |H|^6\) as opposed to \(\lambda_h |H|^6\)). We have also included two operators which are typically eliminated from the Warsaw and SILH bases, namely \(\mathcal{O}_R = |H|^2|D_{\mu}H|^2\) and \(\mathcal{O}_D = |D^2H|^2\) and their associated Wilson coefficients \(c_R\) and \(c_D\). Using integration by parts and equations of motion (equivalently, field redefinitions), these operators can be written in terms of operators in a minimal basis set, \textit{e.g.}~\cite{Giudice:2007fh,Jenkins:2013zja,Elias-Miro:2013eta}.

\begin{table}\renewcommand\arraystretch{1.5}
\centering
{\footnotesize
\begin{tikzpicture}[>=latex]
\begin{scope}[xshift=.7cm, yshift = -3.4cm]
\node[anchor=east] at (-1.05,-.4) {
\scalebox{1.1}{\def\arraystretch{1.8}
\begin{tabular}{|rl|}
  \hline
  \(c_{3G}^{}=\)& \hspace{-3mm}\(\frac{g_s^2}{(4\pi )^2}\frac{1}{20}\) \\
  \(c_{3W}^{}=\)& \hspace{-3mm}\(\frac{g^2}{(4\pi)^2}\frac{1}{20}\)  \\
  \(c_{2G}^{}=\)& \hspace{-3mm}\(\frac{g_s^2}{(4\pi )^2}\frac{1}{20}\) \\
  \(c_{2W}^{}=\)& \hspace{-3mm}\(\frac{g^2}{(4\pi)^2}\frac{1}{20}\)  \\
  \(c_{2B}^{}=\)& \hspace{-3mm}\(\frac{g'^2}{(4\pi)^2}\frac{1}{20}\)  \\
  \hline
\end{tabular}
}
};
\end{scope}
\begin{scope}[xshift=-3.5cm]
\node[anchor=west] at (0,0) {
\scalebox{1.1}{\def\arraystretch{1.8}
\begin{tabular}{|rl|rl|}
  \hline
  \(c_{GG}^{}=\) & \hspace{-3mm}\(\frac{ y_t^2}{(4\pi)^2} \frac{1}{12} \left[ \left( 1 + \frac{1}{12}\frac{g'^2c_{2\b} }{ y_t^2} \right) - \frac{1}{2}\frac{X_t^2}{m_{\tilde{t}}^2} \right]\) &
  \(c_{WB}^{}=\) & \hspace{-3mm}\(-\frac{ y_t^2}{(4\pi )^2}\frac{1}{24}\left[ \left( 1 + \frac{1}{2}\frac{g^2c_{2\b} }{ y_t^2} \right) - \frac{4}{5}\frac{X_t^2}{m_{\tilde{t}}^2} \right]\)  \\
  \(c_{WW}^{}=\) & \hspace{-3mm}\(\frac{ y_t^2}{(4\pi)^2} \frac{1}{16} \left[ \left( 1 - \frac{1}{6}\frac{g'^2c_{2\b} }{ y_t^2} \right) - \frac{2}{5}\frac{X_t^2}{m_{\tilde{t}}^2} \right]\) &
  \(c_W^{}=\) & \hspace{-3mm}\(\frac{ y_t^2}{(4\pi)^2}\frac{1}{40} \frac{X_t^2}{m_{\tilde{t}}^2}\) \\
  \(c_{BB}^{}=\) & \hspace{-3mm}\(\frac{ y_t^2}{(4\pi)^2} \frac{17}{144} \left[ \left( 1 + \frac{31}{102}\frac{g'^2c_{2\b} }{ y_t^2} \right) - \frac{38}{85}\frac{X_t^2}{m_{\tilde{t}}^2} \right]\) &
  \(c_B^{}=\) & \hspace{-3mm}\(\frac{ y_t^2}{(4\pi)^2}\frac{1}{40} \frac{X_t^2}{m_{\tilde{t}}^2}\) \\[.2cm]
  \hline
\end{tabular}
}
};
\end{scope}

\begin{scope}[xshift=.9cm,yshift = -3.4cm]
\node[anchor=west] at (-1.3,-.4) {
\scalebox{1.1}{\def\arraystretch{1.8}
\begin{tabular}{|rl|}
  \hline
  \(c_H^{}=\) & \hspace{-3mm}\(\frac{ y_t^4}{(4\pi)^2}\frac{3}{4}\left[ \left( 1 + \frac{1}{3}\frac{g'^2c_{2\b} }{ y_t^2} + \frac{1}{12} \frac{g'^4c_{2\b}^2}{ y_t^4} \right) - \frac{7}{6}\frac{X_t^2}{m_{\tilde{t}}^2} \left( 1 + \frac{1}{14}\frac{(g^2 + 2g'^2)c_{2\b} }{ y_t^2} \right) + \frac{7}{30}\frac{X_t^4}{m_{\tilde{t}}^2} \right]\) \\
  \(c_T^{}=\) & \hspace{-3mm}\(\frac{ y_t^4}{(4\pi )^2}\frac{1}{4}\left[ \left( 1 + \frac{1}{2}\frac{g^2c_{2\b} }{ y_t^2} \right)^2 - \frac{1}{2}\frac{X_t^2}{m_{\tilde{t}}^2}\left( 1 + \frac{1}{2}\frac{g^2c_{2\b} }{ y_t^2} \right) + \frac{1}{10}\frac{X_t^4}{m_{\tilde{t}}^4} \right]\) \\
  \(c_R^{}=\) & \hspace{-3mm}\(\frac{ y_t^4}{(4\pi)^2}\frac{1}{2}\left[ \left( 1 + \frac{1}{2}\frac{g^2c_{2\b}}{ y_t^2} \right)^2 - \frac{3}{2}\frac{X_t^2}{m_{\tilde{t}}^2}\left( 1 + \frac{1}{12}\frac{(3g^2 + g'^2)c_{2\b}}{ y_t^2} \right) + \frac{3}{10}\frac{X_t^4}{m_{\tilde{t}}^4} \right]\) \\
  \(c_D^{}=\) & \hspace{-3mm}\(\frac{ y_t^2}{(4\pi)^2}\frac{1}{20} \frac{X_t^2}{m_{\tilde{t}}^2}\) \\[0.2cm]
  \hline
\end{tabular}
}
};
\end{scope}

\begin{scope}[xshift=-3.5cm,yshift = -7.cm]
\node[anchor=west] at (0,-.6) {
\scalebox{1.1}{
\begin{tabular}{|rl|}
  \hline
  \(c_6^{}=\) & \hspace{-3mm}\(-\frac{ y_t^6}{(4\pi )^2}\frac{1}{2}\left\{ \def\arraystretch{2.1} \begin{array}{l} \left[ 1 + \frac{1}{12}\frac{(3g^2 - g'^2)c_{2\beta }}{ y_t^2} \right]^3 + \left[ - \frac{1}{12}\frac{(3g^2 + g'^2)c_{2\beta }}{ y_t^2} \right]^3 + \left( 1 + \frac{1}{3}\frac{g'^2c_{2\beta }}{ y_t^2} \right)^3 \\  - \frac{X_t^2}{m_{\tilde{t}}^2}\left[ 2\left( 1 + \frac{1}{12}\frac{(3g^2 - g'^2)c_{2\beta }}{ y_t^2} \right) \left( 1 + \frac{1}{8}\frac{(g^2 + g'^2)c_{2\beta }}{ y_t^2} \right) + \left(1 + \frac{1}{3}\frac{g'^2c_{2\beta }}{ y_t^2} \right)^2 \right] \\+ \frac{X_t^4}{m_{\tilde{t}}^4}\left[1 + \frac{1}{8}\frac{(g^2 + g'^2)c_{2\beta }}{ y_t^2} \right] - \frac{X_t^6}{m_{\tilde{t}}^6}\frac{1}{10} \\ \end{array} \right\}\) \\
  \hline
\end{tabular}
}
};
\end{scope}

\end{tikzpicture}
}
\vspace{-15pt}
\caption{\label{tbl:MSSM}  Wilson coefficients \(c_i\) generated
  from integrating out MSSM stops with degenerate soft mass
  \(m_{\tilde{t}}\),
  and \(\tan \b = \langle
  H_u\rangle / \langle H_d \rangle\) in the MSSM. Table from~\cite{Henning:2014gca}.}
\vspace{-5pt}
\end{table}
\renewcommand\arraystretch{1.0}

\paragraph{2HDM: $H^2S^2,\ (DS)^2$} \textcolor{white}{.}

Here we consider an additional EW scalar doublet \(\Phi\) of hypercharge \(Y_{\Phi} = -1/2\) coupled to the Standard Model, \textit{i.e.} a two Higgs doublet model (2HDM). We impose a \(Z_2\) symmetry on \(\Phi\) (the ``inert'' 2HDM, \textit{e.g.}~\cite{Barbieri:2006dq}), whereupon the most general Lagrangian between \(\Phi\) and the SM is\footnote{Here, \(D_{\mu}\Phi = (\partial_{\mu} - i g W_{\mu}^a \tau^a - ig' Y_{\Phi} B_{\mu})\Phi\), \(\tau^a = \sigma^a/2\) are the \(SU(2)_L\) generators in the fundamental representation, and \(\widetilde{H} \equiv i \sigma_2 H^*\) whence \(\epsilon^{\alpha\beta}\Phi_{\alpha}H_{\beta} = \widetilde{H}^{\dag}\Phi\).}
\begin{align}
\mathcal{L}_{\text{UV}}  &= \mathcal{L}_{\text{SM}} + \left|D_{\mu}\Phi \right|^2 - M^2 \left|\Phi\right|^2 - \frac{\lambda_{\Phi}}{4}\left| \Phi \right|^4 \nonumber\\
&-\lambda_1 | \widetilde{H}|^2 \left|\Phi \right|^2 - \lambda_2 |\widetilde{H}^{\dag}\Phi |^2 -\lambda_3 \big[\big(\widetilde{H}^{\dag}\Phi\big)^2 + \big(\Phi^{\dag}\widetilde{H}\big)^2 \big].
\label{eqn:EW_doub_lag}
\end{align}
The first line contains interactions of the doublet with EW gauge bosons (\(|D\Ph|^2\)), while the second line contains interactions with the Higgs (schematically, \(\Phi^2H^2\)). These interactions are quadratic in \(\Phi\) and therefore lead to one-loop effects.\footnote{In the absence of the \(Z_2\) symmetry taking \(\Phi \to -\Phi\), then the Lagrangian generically contains interactions linear in \(\Phi\) which lead to tree-level effects. Schematically, these take the form \(\Phi H^3\) and Yukawa-like interactions \(\Phi \psi\psi\), which respectively generate \(|H|^6\) and four-fermion operators \(\psi^4\) at tree-level. These were discussed in Sec.~\ref{sec:ESS_SMEFT}; see also~\cite{Henning:2014wua} for further details.}

Integrating out the field \(\Phi\) to one-loop order produces the effective action
\begin{align}
\mathcal{L}_{\text{eff}} = \mathcal{L}_{\text{SM}} &+ \frac{1}{M^2}\bigg(c_6 \mathcal{O}_6 + c_H \mathcal{O}_H + c_T \mathcal{O}_T + c_R \mathcal{O}_R + c_{BB}\mathcal{O}_{BB} + c_{WW}\mathcal{O}_{WW} \nonumber\\ &+ c_{WB}\mathcal{O}_{WB} + c_{3W}\mathcal{O}_{3W} +c_{2W}\mathcal{O}_{2W} + c_{2B}\mathcal{O}_{2B}\bigg) , 
\end{align}
where the Wilson coefficients are given in Table~\ref{tbl:EW_doublet}.
As in the previous example, the same comments about operator normalization and redundancy apply.

\begin{table}
\centering
\[\renewcommand\arraystretch{1.7}
\begin{array}{|rcl|rcl|rcl|}
\hline
c_H &=& \frac{1}{(4\pi)^2} \frac{1}{12} \big( 4\lambda_1^2 + 4 \lambda_1\lambda_2 + \lambda_2^2 + 4\lambda_3^2\big)
& c_{BB} &=& \frac{1}{(4\pi)^2}  \frac{1}{12} Y_\Phi^2(2\lambda_1+\lambda_2)
& c_{3W} &=& \frac{1}{(4\pi)^2}  \frac{1}{60} g^2
\\
c_T &=& \frac{1}{(4\pi)^2}\frac{1}{12}  \big(\lambda_2^2-4\lambda_3^2\big)
& c_{WW} &=& \frac{1}{(4\pi)^2}  \frac{1}{48}  (2\lambda_1+\lambda_2)
& c_{2W} &=& \frac{1}{(4\pi)^2}  \frac{1}{60} g^2
\\
c_R &=& \frac{1}{(4\pi)^2} \frac{1}{6} \big(\lambda_2^2+4\lambda_3^2\big)
& c_{WB} &=& - \frac{1}{(4\pi)^2}  \frac{1}{12} \lambda_2 Y_\Phi
& c_{2B} &=& \frac{1}{(4\pi)^2}  \frac{1}{60} 4g'^2Y_\Phi^2
\\
\hline
\end{array}
\]
\[
\begin{array}{|l|}
\hline
c_6 =- \frac{1}{(4\pi)^2} \frac{1}{6}\Big[ 2\lambda_1^3 + 3\lambda_1^2\lambda_2 + 3\lambda_1\lambda_2^2+\lambda_2^3  + 12\big(\lambda_1+\lambda_2\big)\lambda_3^2\Big]\\
\hline
\end{array}
\]
\vspace{-15pt}
\caption{\label{tbl:EW_doublet}  Wilson coefficients generated from integrating out a massive electroweak scalar doublet \(\Phi\) of hypercharge \(Y_{\Phi} = -1/2\) from the Lagrangian in eq.~\eqref{eqn:EW_doub_lag}. \(g\) and \(g'\) denote the gauge couplings of \(SU(2)_L\) and \(U(1)_Y\), respectively. The couplings \(\lambda_{1,2,3}\) are defined by the Lagrangian in Eq.~\eqref{eqn:EW_doub_lag}. Table adopted from ref.~\cite{Henning:2014wua}.}
\vspace{-5pt}
\end{table}
\renewcommand\arraystretch{1.0}

\paragraph{Neutral vector triplet: $H^2V^2,\ (DV)^2$} \textcolor{white}{.}

As a last example, we consider a specific realization of the neutral vector triplet discussed in Sec.~\ref{sec:unconstrained}. As previously discussed, this model generates tree-level Wilson coefficients. We will show both the tree and one-loop contributions here. Although the focus of this section is on models without tree-level effects, we have included this model as it also shows up as an exeptional case to the discussion in the following subsection~\ref{sec:patterns}.

The model we consider contains a heavy vector in the triplet representation of \(SU(2)_L\) and neutral under \(U(1)_Y\), coupling universally to the SM weak current as discussed in section 2.5.7 of~\cite{Henning:2014wua}.  One way to understand the UV origin of this model is as a deconstructed version~\cite{ArkaniHamed:2001nc} of an extra-dimensional model~\cite{Randall:1999ee} where the EW gauge bosons propagate in the bulk while the SM fermions and Higgs are localized to a boundary. Letting \(Q_{\mu}^a\) denote the heavy vector of mass \(M\), the UV Lagrangian to quadratic order in \(Q\) is given by,
\begin{align}
  \mathcal{L}_{\text{UV}} =& \mathcal{L}_{\text{SM}} + \tilde{g} Q_{\mu}^aJ_{W}^{\mu a} \nonumber \\
  &+ \frac{1}{2} \big(D_{\mu}Q^a_{\nu}D^{\nu}Q^{\mu a} -  D_{\mu}Q_{\nu}^a D^{\mu}Q^{\nu a}  - g\epsilon^{abc} Q_{\mu}^aQ_{\nu}^bW^{\mu \nu c}\big) \nonumber \\
  &+\frac{1}{2}\left(M^2 + \frac{1}{2}\tilde{g}^2|H|^2\right) Q_{\mu}^aQ^{\mu a}.
\end{align}
In the above, the first line contains the tree-level coupling of \(Q_{\mu}^a\) to the EW current \(J_W^{\mu a}\). The second line contains the kinetic term for the vector as well as the magnetic dipole coupling to the EW force (which is universally determined by unitarity~\cite{Weinberg:1970,Ferrara:1992yc}\cite{Henning:2014wua}). The last line contains the mass term and a further coupling to the Higgs field.

Integrating out the heavy vector produces a set of dimension-six operators. At tree-level, we generate \((J_{W}^{\mu a})^2\) which, via the equations of motion \(D^{\nu}W_{\nu \mu}^a = J_{W \mu}^a\), is equivalent to \(\mathcal{O}_{2W}\). At one-loop level, there are further contributions to \(\mathcal{O}_{2W}\) as well as contributions to \(\mathcal{O}_{3W}\), \(\mathcal{O}_H\), and \(\mathcal{O}_6\). Specifically,
\begin{align}
  \Delta \mathcal{L}_{\text{tree}} &= \frac{\tilde{g}^2}{M^2}\mathcal{O}_{2W}, \\
  \Delta \mathcal{L}_{\text{1-loop}} &= \frac{1}{(4\pi)^2}\frac{1}{M^2}\left[\frac{g^2}{20}\big(3\mathcal{O}_{3W} - 37 \mathcal{O}_{2W}\big) + \frac{\tilde{g}^4}{4}\Big( \mathcal{O}_H - \frac{\tilde{g}^2}{6}\mathcal{O}_{6}\Big)\right]. \label{eq:HV_1loop}
\end{align}
Note that there is a consistent limit of this model where \(\tilde{g} \to 0\) while the mass \(M\) and the EW coupling \(g\) are kept fixed.\footnote{In the UV realization given in~\cite{Henning:2014wua}, it corresponds to \(g_2 \to \infty\), \(v \to 0\) and adjusting \(g_1\) to hold \(M\) and \(g\) fixed.} In this limit, the tree-level effects vanish, while at one-loop only modifications to the EW propagators (from \(\mathcal{O}_{2W}\)) and EW triple gauge couplings (from \(\mathcal{O}_{3W}\)) survive. In fact, the ratio of these one-loop effects serves as a powerful discriminant on the spin of the massive particle (in this case, \(c_{2W}^{\text{1-loop}}/c_{3W}^{\text{1-loop}}=-37/3\)), as we discuss in the next subsection.

\subsubsection{Patterns}\label{sec:patterns}

If a new particle is discovered, the two most basic (and most important) questions are (1) what is the particle's mass and (2) what is its spin? These are more-or-less straightforward to answer if the new particle is discovered via direct production. But what if the new particle is ``discovered'' indirectly, by determining that some set of Wilson coefficients in the SM EFT differs from zero? The size of a given Wilson coefficient will clue us in to the mass of the particle, at least up to assumptions on sizes of coupling constants, loop factors, \textit{etc}. But what information might clue us in to the spin of the particle? Well, here comes good news from EW precision tests.

It was recently pointed out in~\cite{Henning:2019vjr} that the ratio \(c_{2W}/c_{3W}\) is very sensitive to the spin of the underlying particle, while remaining largely independent of other details in the microphysics. In more conventional terms, this is the ratio of the EW oblique parameter \(W\) to the anomalous triple gauge coupling (aTGC) \(\lambda_{\gamma}\);\footnote{Recall that the EW oblique parameters \(S,T,U\)~\cite{Peskin:1991sw} and \(V,W,X,Y\)~\cite{Barbieri:2004qk} parameterize deviations from the SM of the 2-point function of EW gauge bosons and the aTGCs~\cite{Hagiwara:1986vm,Hagiwara:1992eh} parameterize deviations from the SM of the coupling among gauge bosons.} to leading order in the EFT
\begin{equation}
  \frac{W}{\lambda_{\gamma}} = -\frac{c_{2W}}{c_{3W}}.
\end{equation}
As we have emphasized, new, heavy particles generically carry EW quantum numbers, and therefore source \(\mathcal{O}_{2W}\) and \(\mathcal{O}_{3W}\).

For simplicity let us assume that the BSM physics is weakly coupled and consists of a single particle of mass \(M\) and charged under \(SU(2)_L\). The generalization to multiple particles is straightforward. Then the precise statement is: barring a single exception, the ratio \(c_{2W}/c_{3W}\) is completely determined by whether the particle is a scalar, fermion, or vector:
\begin{equation}\def\arraystretch{1.2}
  \begin{array}{|c|c|}
  \hline
    & c_{2W}/c_{3W}  \\ \hline
    \text{scalar} & 1 \\
    \text{Dirac fermion} & -4 \\
    \text{vector} & -37/3\\
    \hline
  \end{array}\label{eq:c23W_tab}
  \end{equation}
As is evident this ratio is a pure number---it does not depend, in particuar, on the particle's mass nor on its representation under \(SU(2)_L\) (\textit{i.e.} its EW charge)---and this number varies from \(+1\) for a scalar all the way to \(-37/3 \approx -12\) for a massive vector. Thus, a simple plot of \(c_{2W}\) vs. \(c_{3W}\) gives valuable information on the spin of BSM physics.

Let us explain the origin of the table in eq.~\eqref{eq:c23W_tab}, which will make clear the validity of the table and its simple generalizations. The first point to understand is that any BSM state with EW quantum numbers generates \(\mathcal{O}_{2W,3W}\) at 1-loop, and that its contribution to \(c_{2W,3W}^{\text{1-loop}}\) depends only on the particle's mass, spin, and EW quantum numbers. Our assumption of weakly coupled BSM physics---which concretely means we restrict our attention to renormalizable interactions in the UV theory---further implies that \(\mathcal{O}_{3W}\) is only generated at 1-loop or beyond~\cite{Arzt:1994gp}. The same holds true for \(\mathcal{O}_{2W}\), barring the singular case of the universally coupled neutral vector triplet discussed in sec.~\ref{sec:oneloop_examples}.

The contribution to \(\mathcal{O}_{2W,3W}\) from a particle of mass \(M\) and in the \(R\)th representation of \(SU(2)_L\) were computed in~\cite{Henning:2014wua} where it was found
\begin{equation}
\mathcal{L}_{\text{eff,1-loop}} \supset \frac{1}{(4\pi)^2}\, \frac{1}{M^2}\, \frac{g^2}{60} \, \mu(R) \Big( a_{2\, s} \mathcal{O}_{2W} + a_{3\, s} \mathcal{O}_{3W} \Big) \quad \renewcommand\arraystretch{1.}
\begin{array}{crrrrr}
& a_{2\, s}&  & a_{3\, s}&  \\
\hline
 & 2 & &2 & &\text{scalar}\\
 &  16 & & -4 & &\text{Dirac fermion}\\
 & -37 & & 3 & &\text{vector}\\
\end{array}
\end{equation}
where \(\mu(R)\) is the Dynkin index.\footnote{\(\text{tr}_R(T^aT^b) = \mu(R)\delta^{ab}\). Morally, the index \(\mu(R)\) is a measure of the \(SU(2)_L\) charge; it is the non-abelian analogue of \(Q^2\) for \(U(1)\) groups.} For a real scalar, \(a_{2s} = a_{3s} = 1\). For the massive vector the above is calculated by summing the contributions from the gauge, Goldstone, and ghost fields, as usual. %
We see that in the ratio \(c_{2W}/c_{3W}\) the mass and EW quantum number information drops out. 

If there are multiple particles of different mass or spin or \(SU(2)_L\) charge, then the ratio \(c_{2W}/c_{3W}\) no longer solely depends on the spin. However, as the contributions from the \(i\)th particle scale as \(1/M_i^2\), even a mild hierarchy in mass between BSM states can make \(c_{2W}/c_{3W}\) fairly sensitive to the spin of the lightest state.

As mentioned earlier, there is a singe case where \(\mathcal{O}_{2W}\) can be generated at tree-level: when the UV theory contains a universally coupled neutral vector triplet as discussed in sec.~\ref{sec:oneloop_examples}. However, since \(c_{2W}\) no longer is loop-suppressed, for generic couplings one expects \(c_{2W}/c_{3W} \sim (4\pi)^2\) to be large and therefore a strong indicator about the nature of the BSM physics.

\section{Strongly Coupled BSM}
In this section we discuss strongly coupled dynamics. In this regime, the standard perturbative methods are less efficient and the relations between predictions and fundamental parameters of a given BSM model are fuzzy. Yet, in certain regimes, symmetries and selection rules imply sharp predictions that can be tested: identifying these will be the focus of this section.

\subsection[SMEFT for Composite Higgs Theories]{SMEFT for Composite Higgs Theories\footnote{\bf Based on an original contribution by L.~Vecchi (in review of strongly interacting light Higgs scenarios, see~\cite{Giudice:2007fh,Panico:2015jxa}). }}
\label{sec:CompositeHiggs}

We here derive the SM-EFT under the hypothesis that the Higgs is a resonance of some strong dynamics not far above the TeV scale. These Composite Higgs (CH) scenarios are motivated by the hierarchy problem. We  review the main predictions of Composite Higgs scenarios. This picture singles out two operators as the most important for on-shell Higgs physics, and a small subset of interactions that are relevant for off-shell processes.

We assume that the heavy fields are part of a strongly-coupled sector \emph{characterized by a single mass scale $\Lambda$, and that their self-couplings are all of order $g_*$ }--- that can be weak or strong ($1\lesssim g_*\lesssim4\pi$). The Higgs $H$ is part of the strong dynamics, and therefore couples with full strength $g_*$ to the heavy fields. On the other hand, the SM fermions $\psi$ are allowed to couple with a reduced strength $\epsilon_{\psi}g_*$, with $\epsilon_{\psi}\leq1$; the interactions of the SM gauge bosons follow from gauge invariance. These assumptions are approximately satisfied by all concrete realizations of the CH, and are therefore very reasonable~\cite{Giudice:2007fh}.

Under these assumptions the corrections to the SM Lagrangian can be parameterized as~\cite{Georgi:1992dw}\cite{Cohen:1997rt}\cite{Giudice:2007fh}:
\ba\label{NDA'}
\delta{\cal L}_{\rm NDA}&=&\frac{\Lambda^4}{g_*^2}\hat{\cal L}\left(\frac{g_* H}{\Lambda},\epsilon_{\psi}\frac{g_*\psi}{\Lambda^{3/2}},\frac{D_\mu}{\Lambda}\right),
\ea
where $\hat{\cal L}$ is a function of its arguments and additional model-dependent coefficients of order unity, and $D_\mu$ is a covariant derivative. The small parameters that define the power-counting in this picture are, beyond the SM gauge couplings and $E/\Lambda$, also $g_*v/\Lambda$ and $\epsilon_\psi$.~\footnote{In the Chiral Lagrangian for pions, $g_*v/\Lambda\sim1$ and is not used as an expansion parameter, whereas $\epsilon_\psi$ has no counterpart.}

{In order to evade strong constraints from flavor-violation and other rare processes, the UV dynamics must be approximately {\emph{universal}} in the couplings to the light generations. This may be ensured when the heavy fields couple dominantly to $H$ and the gauge bosons (via gauging), and possibly the third generation, but only weakly to the light SM fermions. In realistic CH models this means that the $\epsilon_\psi$ of the first two generations, see (\ref{NDA'}), are usually negligibly small to be relevant at the LHC. For simplicity, in the following I will assume approximate universality for all generations. In this regime all flavor-violating operators have coefficients controlled by the SM Yukawas. In other words, fermionic operators with powers of $\epsilon_\psi$ and derived from (\ref{NDA'}) approximately satisfy ``minimal flavor violation" (MFV). However, the reader should be warned that this is a mere simplification, as non-universal effects in top and bottom processes are possible in realistic CH models ($\epsilon_{t_L,t_R,b_L,b_R}$ can even be of order unity). This possibility will not be discussed in the following because the associated signatures are more model-dependent than those considered here (see  Refs.~\cite{Pomarol:2008bh,Redi:2011zi,Azatov:2011qy,Panico:2016ull}).}

From (\ref{NDA'}) we obtain \eq{eq:L_SMEFT}, with a certain power-counting for the Wilson coefficients.
In universal theories the most important operators ${\cal O}_i$ appearing in Eq. (\ref{eq:L_SMEFT}) are those that involve the Higgs and gauge fields, see table~\ref{tab1}. The remaining dimension-6 interactions describe non-universal $H$ couplings to fermions, dipole operators, flavor-violating 4-fermion operators. {With present data, those involving the light generations are severely constrained by rare processes and $Z$-pole precision data, and are therefore expected to impact the LHC physics sub-dominantly, see however \cite{Zhang:2016zsp,Franceschini:2017xkh,Grojean:2018dqj}. Non-standard interactions of top and bottom quarks may however be possible, but are not considered here due to our assumption of universality.}


\begin{table}[t]
\begin{center}
\resizebox{\textwidth}{!}
{
\begin{tabular}{|c||c||c|c|c||c|c|} 
\hline
Coefficient &  Generic CH & Light $j=0$ & Light $j=1$ & Only Higgs & $SU(2)$ custodial & NGB-Higgs  \\
\hline
$c_{H,\psi}$ & ${g_*^2}$ & ${g_*^2}$ & ${g_*^2}$ & ${g_*^2}$ & ${g_*^2}$ & ${g_*^2}$  \\
$c_{T}$ & ${g_*^2}$ & ${g_*^2}$ & ${g_*^2}$ & ${g_*^2}$ & ${g_*^2}\times\frac{g'^2}{16\pi^2}$ & ${g_*^2}$  \\
$c_{6}$ & $\frac{g_*^4}{\lambda_h}$ & $\frac{g_*^4}{\lambda_h}$ & $\frac{g_*^4}{\lambda_h}$ & $\frac{g_*^4}{\lambda_h}$ & $\frac{g_*^4}{\lambda_h}$ & $\frac{g_*^4}{\lambda_h}\times\left(\frac{g^2_{\slashed{\cal G}}}{g_*^2}~{\rm or}~\frac{g^2_{\slashed{\cal G}}}{16\pi^2}\right)$ \\
$c_{W,B}$ & $1$ & $\frac{g_*^2}{16\pi^2}$ & $1$ & $\frac{g_*^2}{16\pi^2}$ & $1$ & $1$  \\
$c_{HW,HB}$ & $1$ & $\frac{g_*^2}{16\pi^2}$ & $\frac{g_*^2}{16\pi^2}$ & $\frac{g_*^2}{16\pi^2}$ & $1$ & $1$   \\
$c_{g,\gamma}$ & $1$ & $\frac{g_*^2}{16\pi^2}$ & $\frac{g_*^2}{16\pi^2}$ & $\frac{g_*^2}{16\pi^2}$ & $1$ & $\frac{g^2_{\slashed{\cal G}}}{16\pi^2}$\\
\hline
$c_{2G,2W,2B}$ & $\frac{g_{SM}^2}{g_*^2}$ & $\frac{g_{SM}^2}{g_*^2}\times\frac{g_*^2}{16\pi^2}$ & $\frac{g_{SM}^2}{g_*^2}$ & $\frac{g_{SM}^2}{g_*^2}\times\frac{g_*^2}{16\pi^2}$ & $\frac{g_{SM}^2}{g_*^2}$ & $\frac{g_{SM}^2}{g_*^2}$\\
$c_{3G,3W}$ & $\frac{g_{SM}^2}{g_*^2}$ & $\frac{g_{SM}^2}{g_*^2}\times\frac{g_*^2}{16\pi^2}$ & $\frac{g_{SM}^2}{g_*^2}\times\frac{g_*^2}{16\pi^2}$ & $\frac{g_{SM}^2}{g_*^2}\times\frac{g_*^2}{16\pi^2}$ & $\frac{g_{SM}^2}{g_*^2}$ & $\frac{g_{SM}^2}{g_*^2}$\\
\hline
\end{tabular}
}
\end{center}
\caption{\small Estimate of the EFT coefficients for universal theories characterized by a typical mass scale $\Lambda$ and coupling $g_*$ --- where $g_*\sim g_{\rm SM}$ for weak and $g_*\sim4\pi$ for maximally strong UV completions, for various assumptions (see text). Here $g_{\rm SM}=y_t,g,g',g_s$ denotes a SM coupling. Different combinations of such UV assumptions might simultaneously be satisfied in a given concrete UV model. SM loops and RG evolution from $\Lambda\to\mu<\Lambda$ lead to small corrections to these estimates (or at most comparable) if $g_*\gtrsim g_{\rm SM}$. 
\label{tab2}}
\end{table}

From (\ref{NDA'}) we obtain an estimate for the coefficients $c_i$. The result is shown in the second column in table~\ref{tab2}. {One can make further model-dependent assumptions on the spectrum or the symmetries of the UV dynamics, obtaining coefficients that are suppressed compared to the generic CH case. This is illustrated in the remaining columns.} For example, if the strong dynamics predicts resonances of spin $j$ and mass $\Lambda$, some $c_i$ can be generated at tree-level and thus be enhanced compared to loop-induced ones (of order $g_*^2/16\pi^2$ times smaller).~\footnote{Here we follow~\cite{Giudice:2007fh} and assume that tree-level effects be generated by a UV theory with resonances of spin $j\leq1$ and mass $\Lambda$ that couple via interactions of dimension $\leq4$ and minimally to the gauge bosons. Scalar resonances mixing with the Higgs also generate $\frac{1}{\Lambda^2}(y_\psi\overline{\psi_L}\psi_R)^2$ at tree-level, as well as innocuous renormalizations of the SM parameters. Similarly, at one or higher loops, minimally-coupled universal theories induce operators with fermions that are not included in table~\ref{tab1}. However, their coefficients are dictated by MFV and are expected to impact negligibly our analysis.} The pattern shown in the fifth column is found when the exotic resonances only couple to the Higgs (as in Twin Higgs models). Also, all tree-level contributions can be avoided if the exotic resonances respect a $Z_2$ symmetry that prevents linear couplings to SM currents. Furthermore, a custodial $SU(2)$ can suppress $c_T$ at most by a hyper-charge loop factor. On the other hand, the parameters $c_{6,g,\gamma}$ depend on the nature of the Higgs boson and can be suppressed in scenarios in which the lightness of the Higgs boson is explained promoting it to a Nambu-Goldstone Boson (NGB) of the symmetry breaking pattern ${\cal G}\to{\cal H}$. The small parameter $g_{\slashed{\cal G}}$ and the size of the suppression depend on how the Goldstone symmetry is broken.

\subsubsection{Observable consequences}

The operators in the upper portion of table \ref{tab1} contribute to on-shell Higgs processes~\cite{Contino:2013kra,Elias-Miro:2013mua,Pomarol:2013zra}. The dominant effects are collected in the third column. {The severe constraints from precision electroweak data, especially $Z$-pole physics, force $c_{T}$ and $c_{W}+c_B$ to be very small and thus impact marginally Higgs physics at the LHC (see however \cite{Franceschini:2017xkh,Banerjee:2018bio}).} The operator ${\cal O}_6$ only modifies the Higgs self-couplings and will not be relevant either. The operators ${\cal O}_{H,\psi}$ are singled out as the most important ones for at least two reasons. First, all versions of CH models are expected to generate them ({see first row of} table~\ref{tab2}). Second, their coefficients are usually sizable in strongly-coupled CH scenarios, where $g_*$ can be as large as $4\pi$. They can thus modify all SM Higgs couplings ($h\to VV^*,\psi\bar\psi$) at order $g_*^2v^2/\Lambda^2\approx10\%(g_*/4\pi)^2(10~{\rm TeV}/\Lambda)^2$, that is significant even if the UV physics is very heavy. Via loops, the operators ${\cal O}_{H,\psi}$ also affect the SM predictions for $h\to gg,\gamma\gamma,\gamma Z$ at the same order. ${\cal O}_{HW,HB,g,\gamma}$ dominantly modify radiative Higgs decays and may also be important; however, their relevance is more model-dependent.

The operators in table~\ref{tab1} can also be constrained by taking advantage of the growth in energy appearing in processes at high momentum transfer $q^2\gg m^2_W$:
\ba\label{A}
{\cal A}={\cal A}_{\rm SM}+{\cal A}_{\rm BSM}\times{c}\frac{q^2}{\Lambda^2}+{\cal O}\left({c}'\frac{q^4}{\Lambda^4}\right)\,,
\ea
where $c$ and $c^\prime$ are dimensionless (combinations of) Wilson coefficients.
Familiar examples are vector boson scattering $V_LV_L\to V_LV_L, hh$ and $V_LV_L\to t\bar t$. Other processes are listed in the last column in the table. 

As emphasized in sec.~\ref{sec:CompositeHiggs}, the assumptions on the UV dynamics are essential in extracting reliable constraints on the new physics parameters~\cite{Contino:2016jqw}. Without hypothesis on the UV physics we do not know the size of $c$ in (\ref{A}) and cannot translate the experimental bound on ${\cal A}$ into a constraint on the parameters $\Lambda,g_*$. Similarly, we cannot estimate the error incurred in neglecting ${c}'{q^4}/{\Lambda^4}$ if we do not know $c'/c$. This is precisely the information provided by the power-counting {discussed in sec.~\ref{sec:CompositeHiggs}}. 
In CH scenarios this crucial input is summarized in table~\ref{tab2}. Once a power-counting is assumed, only a very small number of operators turns out to be relevant for a given observable.
This is what makes the SM-EFT a useful tool. For example, inspecting table~\ref{tab2} one easily sees that Drell-Yan processes are controlled by ${\cal O}_{2W,2B}$~\cite{Farina:2016rws}, whereas $pp\to jj$ are dominated by ${\cal O}_{2G}$~\cite{Alioli:2017jdo}. Similar considerations apply to di-boson events. In this case 4 operators are expected to dominate, ${\cal O}_{W,B,HW,HB}$, but in reality only two linear combinations turn out to be relevant~\cite{Franceschini:2017xkh}.

\subsection[Composite Vectors and Multipolar Interactions]{Composite Vectors and Multipolar Interactions\footnote{\bf Based on a contribution by D.~Liu and I.~Low, contains material inspired by Ref.~\cite{Liu:2016idz}.}}\label{sec:remedios}
In this section we discuss the possibility that the transverse polarisations of vector bosons be strongly interacting. This seems in apparent contrast with the weakly coupled nature of the SM gauge bosons. Ref.~\cite{Liu:2016idz} however proposes a structure for which the SM and its deviations are not necessarily associated with the same coupling strength. This is interesting in the context of searches for anomalous triple gauge couplings (TGCs): the models discussed here might induce large deviations there, or in processes involving Higgs bosons and vectors, such as in the decay $h\to Z\gamma$. In what follows we assume that the new physics preserves custodial symmetry such that $c_T = 0$. 

It is useful to first recall the power-counting rules associated with the CH models discussed in the previous section. There are two basic assumptions: one is that the  Higgs  and the longitudinal components of the SM gauge bosons are pseudo-Nambu-Goldstone-bosons associated with the global symmetry breaking of a strong sector; the other is that SM fermions acquire masses from their linear mixing with the strong sector. This leads to the following power-counting rules in the Lagrangian:

\begin{itemize}
\item Higgs fields will be associated with a strong coupling $g_*$ in the operators which preserve the symmetry of the strong sector including the non-linearized symmetry.
\item Explicitly breaking of the strong sector symmetry will be associated with SM gauge couplings and Yukawa couplings $g,g^\prime, y_f$.
\end{itemize}
Following these rules, we have summarized the size of the Wilson coefficients of the operators for the SILH scenario in Table~\ref{tab:powercounting}. We see that only $\mO_H, \mO_\psi$ are enhanced by strong coupling, which make them the most relevant operators in the SILH scenario, as discussed in section~\ref{sec:CompositeHiggs}.

Ref.~\cite{Liu:2016idz} considers the possibility that the SM transverse gauge bosons are part of the strong dynamics. These scenario was dubbed \emph{Remedios} and is based on the observation that the normal SM gauge interactions (mono-pole) and multi-pole interactions (involving the field strength and its derivatives) have  different symmetry structure. Therefore, they can have different coupling strength in principle. The small  coupling $g$ controls the renormalizable interactions between the gauge boson and the fermions, while the large coupling $g_*$ determines the multi-pole interaction with the resonances of the strong sector. This will lead to new power-counting rules for the gauge bosons:
\begin{itemize}
\item The field strengths of the gauge boson and their derivatives will be associated with a strong coupling $g_*$, while the normal gauge interactions are realized by  changing the partial derivative to covariant derivative: $\partial_\mu \rightarrow D_\mu = \partial_\mu -i  gA_\mu$.
\end{itemize}
This leads to the case that the $\mO_{3W}$ is enhanced by the strong coupling and the $\mO_{2W,2B}$ have $\mO(1)$ Wilson coefficient, which has been summarized in the third row of Table~\ref{tab:powercounting}.
We can also consider the scenario that both transverse gauge bosons and Higgs bosons are part of the strong dynamics. This is possible if the nonlinearly realized symmetries of the Higgs sector follow specific patterns:

\begin{itemize}
\item \text{Remedios + minimal CH models (MCHM):}: the symmetry breaking of the strong sector will be $SO(5)\times \widetilde{SU(2)} \times U(1)_X \rightarrow SO(4)\times \widetilde{SU(2)} \times U(1)_X$, where another global symmetry $\widetilde{SU(2)}$ is needed to stablize the Higgs potential. 
\item \text{Remedios} +$ ISO(4)$: the symmetry breaking of the strong sector will be $ISO(4)\times  U(1)_X \rightarrow SO(4) \times U(1)_X$, where the $ISO(4)$ is the non-compact group $SO(4)\rtimes T^4$.
\end{itemize}
The corresponding power-counting rules for the size of the Wilson coefficients are presented in the fourth and fifth rows of Table~\ref{tab:powercounting}. We see that $\mO_{HW}$ can be enhanced by the strong coupling $g_*$ in the second case.

 \begin{table}[t]
\renewcommand{\arraystretch}{1.2}
{
\begin{center}
\begin{tabular}{|c||c|c|c|c|c|c|c|c|c|c|}
\hline 
Model & $ \mO_H$ &${\cal O}_{2W}$ & ${\cal O}_{2B}$& ${\cal O}_{3W}$ & ${\cal O}_{HW}$ & ${\cal O}_{HB}$ & ${\cal O}_{W,B}$ & ${\cal O}_{BB}$ & $\mO_\psi$  \\
\hline
{\footnotesize SILH (see sec. ~\ref{sec:CompositeHiggs})}& ${g_*^2}$    & $\frac{g^2}{g_*^2}$  & $\frac{g^{\prime2}}{g_*^2}$ &$\frac{g^2}{16\pi^2}$& $\frac{g^2_*}{16\pi^2}$   & $\frac{g^2_*}{16\pi^2}$&   $1$ &$\frac{g^2}{16\pi^2}$ & $\frac{g_*^2}{g^2}$     \\
{\footnotesize Remedios }&  & 1 & 1& $\frac{g_*}{g}$&  &  &  &   &   \\
{\footnotesize  Remedios+MCHM  } &${g_*^2}$    & 1&  1&$\frac{g_*}{g}$& $1$       & $1$ &
$1$ & 1 &  ${g_*^2}$  \\
{\footnotesize  Remedios+$ISO(4)$  }  &${\lambda_h}$    &1 & 1& $\frac{g_*}{g}$ & $\frac{g_*}{g}$ & $1$   &  $1$   & 1 &  ${\lambda_h}$  \\
\hline
\end{tabular}
\end{center}
}
\vspace{0.3cm}
 \caption{Size of the Wilson coefficients in different scenarios, where $g_*$ denotes a strong coupling in the new physics sector.}
 \label{tab:powercounting}
\end{table}

\section{Conclusions}

In this document we collected several \emph{benchmark} scenarios for both weakly and strongly coupled new physics. For each of them, a mapping to the relevant dimension-6 operators in the SM EFT, as well as the expected size of the coefficients, is shown in tables \ref{tab:bottom-up},\ref{tab:dim6ops_note},\ref{tbl:MSSM},\ref{tbl:EW_doublet},\ref{tab2},\ref{tab:powercounting}. Furthermore, for several Higgs and electroweak processes we show table~\ref{tab1} what are the most relevant operators (in universal theories).

On the one hand, this table together with the maps from models to EFT coefficients, can help experimentalists designing an analysis for a given process to individuate the relevant small set of operators, motivated by some concrete BSM scenario. This can be used to justify a choice of a specific set of operators.
On the other hand, the same tables can also be used to easily translate the constraints on EFT coefficients to limits in the parameter space of one of the BSM models studied here.

As a long-term goal, a global SM EFT analysis of all available experimental low and high-energy data, without assumptions on the EFT coefficients, is certainly worthy the present effort which is being put into it.
Nevertheless, if one wants to learn about some more specific class of BSM models, it is reasonable to consider only the limited set of operators generated in that scenario with largest coefficients. Such restrictions in parameter space often makes the limits much stronger, thus improving our knowledge on those new physics models.
The BSM benchmarks presented here cover a wide range of possible new physics, from weakly-coupled simplified models, to strongly coupled theories.

\subsection*{Acknowledgments}

We are grateful to all members of the LHC Higgs WG for stimulating discussions that led to this document. We thank in particular A.~Gritsan, C.~Grojean and V.~Sanz, for important comments on the manuscript. The work of  D.~Liu and I.~Low was supported by the United States Department of Energy under Grant Contract DE-SC0012704.

\bibliography{Biblio}{}
\bibliographystyle{JHEP}

\end{document}